\documentclass[aps,prb,longbibliography,twocolumn,floatfix,superscriptaddress,nofootinbib]{revtex4-2}

\usepackage{graphicx}
\usepackage{amsmath}
\usepackage{amssymb}
\usepackage{tikz}
\usepackage{amsfonts}
\usepackage{physics}
\usepackage{xcolor} 
\usepackage[colorlinks=true, linkcolor=blue, citecolor=blue, urlcolor=blue]{hyperref}
\usetikzlibrary{shapes,shadows}
\usetikzlibrary{calc}
\usetikzlibrary{backgrounds}

\definecolor{darkblue}{rgb}{0,0,.65}
\definecolor{darkgreen}{rgb}{0.3,0.6,0.3}
\definecolor{darkorange}{rgb}{0.85,0.65,0.3}
\definecolor{cyan1}{rgb}{0.0, 0.6, 0.6}

\newcommand{\imk}[1]{{{#1}}}
\newcommand{\mk}[1]{{{#1}}}

\usepackage{soul}
\usepackage{ulem} 

\begin{document}

\title{Limitations of Avila-inspired theory in the diamond chain with quasiperiodic disorder}

\author{Manish Kumar}
\email{manish22@iiserb.ac.in}
\affiliation{Department of Physics, Indian Institute of Science Education and Research, Bhopal, Madhya Pradesh 462066, India}
\author{Ivan M. Khaymovich}
\email{ivan.khaymovich@gmail.com}
\affiliation{Univ. Toulouse, CNRS, Laboratoire de Physique Th{\'e}orique, Toulouse, France}

\author{Auditya Sharma}
 \email{auditya@iiserb.ac.in}
\affiliation{Department of Physics, Indian Institute of Science Education and Research, Bhopal, Madhya Pradesh 462066, India}
\date{\today}

\begin{abstract}
  The mobility edges (MEs) that separate localized, multifractal and
  ergodic states in energy are a central concept in understanding
  Anderson localization. In this work we study the effect of several
  mutually commensurate quasiperiodic frequencies on the mobility edge
  formation. We focus on the example of the addition of a constant
  offset to the quasiperiodic potential of the one-dimensional
  all-bands-flat diamond chain. We show that this additional offset
  can transform the anomalous mobility edges (AMEs), i.e. the
  energies, separating localized and multifractal states, into
  conventional mobility edges, separating localized from ergodic
  states. We further compare the numerically obtained ME
structure with an analytical estimate motivated by Avila's global
theory. While the estimate provides useful insight into the mobility edge
formation, discrepancies with the numerical results appear both at a
quantitative level, through shifts in the mobility edge locations, and at a
qualitative level, through the emergence of multiple mobility edges not
captured by the analytical estimate.
\end{abstract}

\maketitle
\section{Introduction}

Anderson localization is a paradigmatic phenomenon in condensed matter
physics, in which on-site disorder causes eigenstates to become
localized~\cite{PhysRev.109.1492}. Unlike random disorder where all
states localize in one dimension in the presence of the tiniest of
disorder, quasiperiodic potentials can exhibit a metal-to-insulator
transition even in one dimension. The Aubry-Andr{\'e}-Harper (AAH)
model, which undergoes a transition from a fully delocalized phase to a
fully localized phase as the strength of the quasiperiodic potential
increases~\cite{aubry1980} is a prototype of this
phenomenon. Furthermore, models with quasiperiodic potentials can host
both localized and delocalized states separated by mobility edges
(MEs)~\cite{PhysRevLett.61.2144, Y_Hashimoto_1992, PhysRevA.75.063404,
  PhysRevA.80.021603, PhysRevLett.104.070601, PhysRevA.90.061602,
  PhysRevB.96.085119, PhysRevLett.123.070405, PhysRevLett.114.146601,
  Xu_2020, PhysRevB.101.174205, PhysRevResearch.2.033052}.  The
self-duality of the standard AAH model eliminates the mobility edge,
but modified versions featuring long-range
hopping~\cite{PhysRevLett.104.070601, PhysRevLett.123.025301,
  PhysRevB.100.174201, PhysRevB.103.075124}, broken
duality~\cite{PhysRevB.96.085119, PhysRevLett.123.070405,
  PhysRevLett.124.113601}, diluted
potentials~\cite{PhysRevLett.125.196604, PhysRevLett.131.176401}, or
spin-orbit coupling~\cite{PhysRevA.87.023625, PhysRevB.77.134204}—can
exhibit MEs. Systems featuring MEs can show enhanced thermoelectric
performance, offering promising potential for thermoelectric device
applications~\cite{PhysRevLett.112.130601, PhysRevB.96.155201,
  PhysRevResearch.2.013093}. The Aubry-Andr{\'e} (AA) potential has been experimentally
realized using ultracold atoms to study both
single-particle~\cite{PhysRevLett.103.013901, PhysRevLett.95.070401,
  PhysRevLett.106.230403} and many-body
localization~\cite{science.aaa7432}, sparking renewed interest in
quasiperiodic systems at zero~\cite{PhysRevB.75.155111, CRPHYS2018,
  RevModPhys.91.021001, PhysRevB.87.134202, PhysRevLett.115.230401,
  pnas.1800589115, PhysRevResearch.1.032039} and
finite~\cite{PhysRevA.84.063614, Roy_2018, PhysRevLett.113.045304}
temperatures.

\begin{figure}
    \centering
    \includegraphics[width=0.4\textwidth]{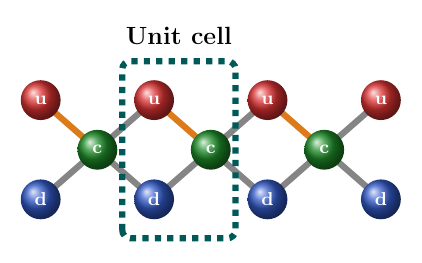}
\caption{Schematic of diamond lattice with $u$ (top), $c$ (center), and $d$ (bottom) sites. Solid grey (orange) lines between lattice sites have the same hopping amplitude and negative (positive) sign. Dashed rectangle shows the unit cell containing three sites.}

    \label{Diamond_chain_schematic}
\end{figure}

In this work, we investigate the emergence of conventional mobility
edges and anomalous mobility edges (AMEs) in a flat
band~\cite{CRPHYS2013, S021797921330017X, S0217979215300078,
  Flach_2014, Leykam01012018, PhysRevLett.85.3906, PhysRevLett.88.227005, Gladchenko2009, PhysRevLett.81.5888, PhysRevB.64.155306} system also known as the all-bands-flat (ABF)
diamond chain~\cite{nilanjan}, which supports compact localized states
(CLS)~\cite{PhysRevB.95.115135, Sathe_2021} as exact eigenstates in
the zero disorder limit, with the spectrum given by $3$ degenerate flat bands. In the presence of weak uniform random
disorder, the ABF diamond chain exhibits flat band based localization
that remains extremely weak~\cite{nilanjan}. When this disorder is
replaced by a quasiperiodic Aubry-Andr{\'e} potential, the
eigenstates become
multifractal, a subtle and increasingly studied feature of quantum
systems. In the ABF diamond chain, each unit cell consists of three
sites labeled by $u_n$ (up), $c_n$ (center), and $d_n$ (down) as shown
in Fig.~\ref{Diamond_chain_schematic}. Interestingly, applying a
symmetric quasiperiodic potential, where the same on-site potential
is assigned to both the top and bottom sites of each unit
cell, completely lifts the degeneracy of the flat bands while preserving the
robustness of the CLS~\cite{PhysRevB.91.235134}. In contrast, an
antisymmetric AA potential, where the on-site potentials at the top
and bottom sites of each unit cell are equal in magnitude but opposite
in sign and zero at the central sites, destroys both degeneracy of the flat bands and compact localization. Under
such antisymmetric quasiperiodic modulation, all the states in this
model are multifractal below a critical value of on-site
disorder. Above this critical value AMEs, separating multifractal and
localized states, emerge~\cite{aamna}.

\begin{figure}
    \centering
    \includegraphics[width=0.48\textwidth]{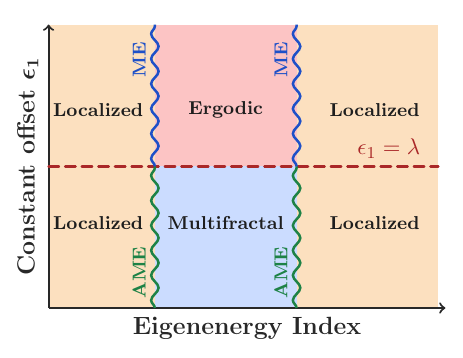}
\caption{A schematic showing the presence of AMEs in
  $\epsilon_1<\lambda$ regime and MEs in $\epsilon_1>\lambda$
  regime. The boundaries represented by wiggly lines are shown
    only for illustrative purposes and do not represent the actual
    dependence of the mobility edge positions on eigenstate index.}

    \label{main_results}
\end{figure}

We introduce a constant offset $\epsilon_1>0$ to the Aubry-Andr{\'e} harmonic potential of amplitude $\lambda$ and apply it
antisymmetrically across the ABF diamond chain. This modification
leads to the appearance of several quasiperiodic contributions with
commensurate frequencies in the effective one-dimensional
Hamiltonian and, in turn, to unexpected and nontrivial changes in the
nature of the eigenstates. Specifically, we observe the emergence of
AMEs in the regime $\epsilon_1 < \lambda$. These AMEs originate from
singularities in the transfer matrix due to vanishing effective hopping
amplitudes of the above 1D Hamiltonian, accompanied by vanishingly
small Lyapunov exponents (LEs)~\cite{PhysRevLett.131.176401}. In the
other limit, $\epsilon_1 > \lambda$, we find conventional MEs that
separate ergodic and localized states. Thus, we present a model,
having both anomalous and conventional MEs in different parameter
ranges (Fig.~\ref{main_results}), a feature not reported in earlier studies. A common
diagnostic for the characterization of localization properties of
eigenstates $\psi_n(E)$ is the energy $E$-resolved inverse participation ratio $I_q = \sum_{n=1}^{N}
|\psi_n(E)|^{2q} \sim N^{(1-q)D_q}$, where $n$ is a spatial index, and $D_q$ is known as the
fractal dimension. In our analysis, we use fractal dimension $D_2$
corresponding to $q=2$, to characterise the localization properties of
eigenstates. We further confirm these results by calculating the
Lyapunov exponent $\gamma$ which is positive for localized states
and $0$ for delocalized, both ergodic and multifractal states.

In recent years, obtaining analytical expressions for mobility edges
in various systems has received a great deal of interest. A major
breakthrough in this direction has been the use of Avila's global
theory~\cite{Avila2015}, one of the seminal contributions that earned
him the Fields Medal. This theory has since been applied widely, and
its predictions have been verified numerically across multiple
models~\cite{PhysRevLett.125.196604, PhysRevB.108.174202,
  PhysRevB.105.174206, Cai_2023, PhysRevB.103.174205}. In the case of
the ABF diamond chain with the AA potential but without any constant
offset, the analytical expression for the ME ($|E|=4/\lambda$) has
already been obtained~\cite{aamna,PhysRevB.107.014204} using insights
from the extended Harper model~\cite{Avila2017}. In this work, we apply an Avila-inspired analytical approach to the ABF diamond
chain with both the AA potential and a constant offset $\epsilon_1\ge0$.
We find that the resulting analytical estimate does not provide a
complete description of the mobility edge structure in this model.
As a matter of fact we find that it only manages
  to separate localized states of one kind from localized states of a
  different kind. But surprisingly, the expression obtained from the Avila-inspired analysis
agrees with the extended Harper model result when $\epsilon_1 =
0$. As a result of the above limitation, for $\epsilon_1 > 0$,
the analytical \imk{estimate} from Avila-inspired analysis remains only
qualitatively valid for $\lambda \gtrsim 2$ and fails
quantitatively. To achieve quantitative agreement, we introduce
fitting parameters into the analytical formula and verify the modified
expression through careful numerical analysis for several values of
$\lambda \gtrsim 2$. Furthermore, we demonstrate for $\lambda
  \lesssim 2$, the emergence of multiple mobility edges that lie beyond the scope of the analytical estimate.

The paper is organized as follows: In Section~\ref{sec:Model} we
introduce the model. In Section~\ref{sec:analytical} we apply an Avila-inspired analytical approach to our model and derive an estimate for the mobility edge structure. Section~\ref{sec:D2} is devoted to multifractal
analysis to determine ME locations numerically. We confirm the results
by calculating the Lyapunov exponent in Sec.~\ref{sec:Lyapunov Exponent}. Finally, we discuss the
  discrepancy between our numerical results and the analytical estimate motivated by Avila's global theory and heuristic fitting methods for obtaining
  approximate expressions of MEs in section
~\ref{sec:Avila_discrepancy}. The results are summarized in
Sec.~\ref{sec:summary}.  An additional phase diagram for $\lambda =
0.8$ is provided in Appendix~\ref{app:lambda=0.8}. Numerical results
are supported by Lyapunov exponent analysis in
Appendix~\ref{app:Lyapunov_exp_vs_N}. Appendix~\ref{app:LE_4.4_multiple}
demonstrates the emergence of multiple anomalous mobility edges at
higher potential strength. We further confirm our results with $I_2$
collapse analysis in Appendix~\ref{app:IPR_collapse}. Finally,
Appendix~\ref{app:single_realization} presents results for a single
realization of the phase parameter $\theta$.

\section{Model}\label{sec:Model}
 We consider a single-particle model on a quasi-one-dimensional
 diamond lattice (see Fig.~\ref{Diamond_chain_schematic}), whose Hamiltonian is given by ~\cite{aamna,nilanjan}:
\begin{equation}
  \hat{H}=\hat{H}_{\text{hop}}+\hat{H}_{\text{os}},
  \label{eq:hamiltonian}
\end{equation}
where
\begin{equation}\label{eq:ham}
\begin{aligned}
\hat{H}_{\text {hop }}= & -J \sum_{n=1}^{N/3}\left(-\hat{u}_n^{\dagger} \hat{c}_n+\hat{c}_n^{\dagger} \hat{d}_n+\hat{c}_n^{\dagger} \hat{u}_{n+1}+\hat{c}_n^{\dagger} \hat{d}_{n+1}\right)+\text { H.c., } \\
\hat{H}_{\mathrm{os}}= & \sum_{n=1}^{N / 3}\left(\zeta_n^u \hat{u}_n^{\dagger} \hat{u}_n+\zeta_n^c \hat{c}_n^{\dagger} \hat{c}_n+\zeta_n^d \hat{d}_n^{\dagger} \hat{d}_n\right).
\end{aligned}
\end{equation}
Here, each unit cell $n$ consists of three sites, hence the total
number of sites denoted by $N=3L$ should be a multiple of
3. $\hat{u}_n^{\dagger}$, $\hat{c}_n^{\dagger}$, and
$\hat{d}_n^{\dagger}$ are fermionic creation operators acting at the u
(up), c (center), and d (down) sites respectively in the $n^{th}$ unit
cell as shown in the schematic
(Fig.~\ref{Diamond_chain_schematic}). $J$ is the nearest neighbor
hopping amplitude, which we choose to be $J=1$ without loss of generality.  $\zeta_n^\alpha$ ($\alpha$=u,c,d) denotes on-site
potential on the $n^{\text{th}}$ site. The on-site energies are drawn from a
quasiperiodic antisymmetric potential according to:
\begin{equation}
     \zeta_n^c = 0, \quad \zeta_n^u = -\zeta_n^d= \epsilon_1+\lambda \cos[2\pi\beta n+\theta]
     \label{antisymmetric_potential}
\end{equation}
where $\epsilon_1$ is an offset potential, and $\lambda$ is the potential
strength. The quasiperiodicity parameter $\beta$
is an irrational number to ensure the incommensurate nature of the
potential. In order to implement periodic boundary conditions, we take
the number of unit cells to be $L=F_l$ (the $l$th Fibonacci number), and
adopt the rational approximant $\beta_l \equiv F_{l-1}/F_l$ of
$\beta = (\sqrt{5}-1)/2$. In the limit $l\to\infty$, $\beta_l$
converges to the inverse golden ratio $\beta$, thereby preserving
quasiperiodicity. $\theta$ is chosen randomly from a uniform
distribution between $[0,2\pi]$. Unless explicitly stated otherwise, all quantities are averaged over 50 realizations of $\theta$.

In the absence of disorder ($\zeta_n^\alpha=0$), the diamond chain
exhibits three flat bands at energies $\pm 2$, 0 and for this reason
this system is also known as the all-bands-flat diamond chain. Also,
this system possesses only compactly localized states. Hence, one can
conclude that in the zero disorder limit, the system is highly
degenerate, a good insulator and possesses chiral symmetry. If we put
a symmetric potential, where the same potential is applied to the top
and bottom sites of a unit cell, $\zeta_n^u = \zeta_n^d$, compact localization is preserved
although degeneracy is lifted. In this work we consider the
quasiperiodic potential applied in an antisymmetric manner (see \eqref{antisymmetric_potential}). For $\epsilon_1=0$, this model reduces to the case which has been
already studied. In the antisymmetric case, not only is
the degeneracy lifted but the compact localization of eigenstates is
also destroyed. In the low disorder regime, all the eigenstates are
multifractal while with increasing disorder strength
$\lambda$, anomalous mobility edges, which separate localized and
multifractal states, are observed~\cite{aamna}.

\section{Avila's Global theory} \label{sec:analytical}

In this section, we demonstrate how an analytical estimate of the
MEs may be obtained using an Avila-inspired approach~\cite{Avila2015}. We start by computing the
transfer matrix of our system. To write the transfer matrix, we first
write the lattice equation of the effective 1D model of the diamond
chain. Starting from the Schr\"{o}dinger equation of the Hamiltonian in the site basis, we obtain the following set of coupled equations for the
components $u_k$, $d_k$, and $c_k$ of the wavefunction $\psi_k(E) = (u_k,
d_k, c_k)^T$ in the $k$-th unit cell:
\begin{align}
E u_k &= \zeta_k u_k + J c_k - J c_{k-1}, \nonumber \\
E c_k &= J u_k - J d_k - J u_{k+1} - J d_{k+1},\nonumber \\
E d_k &= -\zeta_k d_k - J c_k - J c_{k-1}.
\end{align}
To simplify the structure of the equations, we perform the unitary transformation
\begin{align}
p_k &= \frac{u_k - d_k}{2} + \frac{c_k}{\sqrt{2}}, \nonumber \\
q_k &= -\frac{u_k - d_k}{2} + \frac{c_k}{\sqrt{2}}, \nonumber \\
s_k &= \frac{u_k + d_k}{\sqrt{2}},
\end{align}
which diagonalizes the intracell hopping matrix and renders the antisymmetric structure explicit. In the case of antisymmetric on-site disorder the Schr\"{o}dinger equation in the rotated basis $(p_k,q_k, s_k)$ takes the form
\begin{align}
E p_k &= \frac{\zeta_k}{\sqrt{2}} s_k - \big(-\sqrt{2} p_k + s_{k+1}\big), \label{eq:p_comp} \\
E q_k &= -\frac{\zeta_k}{\sqrt{2}} s_k - \big(\sqrt{2} q_k + s_{k+1}\big), \label{eq:q_comp} \\
E s_k &= \frac{\zeta_k}{\sqrt{2}} (p_k - q_k) - (p_{k-1} + q_{k-1}). \label{eq:s_comp}
\end{align}
Substituting the expressions for $p_k$ and $q_k$ from the first two
equations into the third expression, we obtain the lattice equation for
the diamond chain in the case of antisymmetric on-site disorder:
\begin{equation}
   E[E^2-4]s_k=\zeta_k^2Es_k-2\zeta_ks_{k+1}-2\zeta_{k-1}s_{k-1},
   \label{Lattice_equation}
\end{equation}
with $\zeta_k=\epsilon_1+\lambda \cos [2\pi k\beta+\theta]$, the effective energy $E(E^2-4)$, quasiperiodic $E$-dependent effective on-site potential  $\zeta_k^2 E$ with a period $\pi/\beta$, and the quasiperiodically modulated hopping $2\zeta_k$, with another period $2\pi/\beta$. One can write this equation in terms of the transfer matrix also as
\begin{equation}
\Phi_{k+1}=T_k \Phi_k
\end{equation}
where $\Phi_k^T=(s_{k} \hspace{2mm} s_{k-1})^T$. $T_k$ is the transfer matrix:
\begin{equation}
  T_k(\theta)=\begin{pmatrix}
    \frac{E}{2}\left[\zeta_k-\frac{E^2-4}{\zeta_k}\right] & -\frac{\zeta_{k-1}}{\zeta_k}\\
    1 & 0
  \end{pmatrix}
  \label{transfer_matrix}
\end{equation}
The next
step is to calculate the Lyapunov exponent which is related to
localization length. The Lyapunov exponent is inversely related to the
localization length, hence, for a delocalized eigenstate it decreases
with increasing system size and tends to zero in the thermodynamic
limit. In contrast, localized states are characterized by a finite
LE. The transfer matrix can be used to calculate the
LE:
\begin{equation}
  \gamma(E)=\lim_{L\rightarrow\infty}\frac{1}{L}\ln\frac{|\Phi_L|}{|\Phi_0|},
\end{equation}
where $L$ is the number of unit cells. The transfer-matrix state can be written as
\begin{equation}
  \Phi_L=T_LT_{L-1} \cdots T_2T_1\Phi_0
\end{equation}
where $T_L$ is the transfer matrix. Putting it in $\gamma(E)$, we get
\begin{equation}
    \gamma(E) =\lim_{L\rightarrow\infty}\frac{1}{L}\ln\frac{|T_L T_{L-1} \cdots T_2 T_1 \Phi_0|}{|\Phi_0|}.
    \end{equation}
For almost any choice of $\Phi_0$, the growth rate of $\Phi_i$ is governed by the same Lyapunov exponent~\cite{PhysRevB.52.4146}. This leads to the simplification:
\begin{align}
    \gamma(E) &= \lim_{L\rightarrow\infty} \frac{1}{L}\ln\left\| T_L T_{L-1}\cdots T_1 \right\|
  \nonumber \\
              &=   \lim_{L \rightarrow \infty} \frac{1}{L} \ln \left\lVert \prod\limits_{k=1}^L T_k \right\rVert.
              \label{gamma_sum}
\end{align}
In the transfer matrix~\eqref{transfer_matrix}, we can't use Avila's
theory directly as $\zeta_k$ has incommensurately distributed zeros. So we first
pull out the singularity as a factor by writing~\cite{singularity_remove}:
\begin{align}
  T_k(\theta) &=\frac{1}{2\zeta_k}\begin{pmatrix}
    E\left[\zeta_k^2-E^2+4\right] & -2\zeta_{k-1}\\
    2\zeta_k & 0
  \end{pmatrix} \nonumber \\
     &= A_kB_k
\end{align}
where
\begin{align}
 A_k(\theta) =\frac{1}{2\zeta_k}, \quad
 B_k(\theta) = \begin{pmatrix}
    E\left[\zeta_k^2-E^2+4\right] & -2\zeta_{k-1}\\
    2\zeta_k & 0
  \end{pmatrix}.
  \label{T_decompose}
\end{align}
Since $T_k(\theta)$ is decomposed into two parts as in Eq.~\eqref{T_decompose}, \mk{hence to calculate the total LE $\gamma(E)$, we evaluate $\gamma^A(E)$ and $\gamma^B(E)$ separately}. For $\gamma^A(E)$ we have
\begin{align}
 \gamma^A(E)&=  \lim_{L \rightarrow \infty} \frac{1}{L} \ln \left\lVert \prod\limits_{k=1}^L A_k(\theta) \right\rVert \nonumber \\
       &= \lim_{L \rightarrow \infty} \frac{1}{L} \ln \left\lVert \prod\limits_{k=1}^L \frac{1}{2(\epsilon_1+\lambda \cos [2\pi k\beta+\theta])} \right\rVert.
\end{align}
Since $\beta$ is an irrational number, as $k$ varies, the argument of the cosine fills the interval $(0,2\pi]$ uniformly. This follows from Weyl's equidistribution
  theorem and properties of irrational rotations~\cite{Weyl1916,
    irrational}. Then by using the classical Jensen's
  formula~\cite{PhysRevB.100.125157, gradshteyn_ryzhik_2007} and
  ergodic theory, we can write
\begin{align}
\gamma^A(E)&= \frac{1}{2\pi}\int_0^{2\pi} \ln \frac{1}{2(\epsilon_1+\lambda \cos \theta)} d\theta \nonumber \\
          &=
        \begin{cases}
            \ln \frac{1}{\lambda} & \text{if } \epsilon_1 < \lambda \quad \\
            \ln \frac{1}{\epsilon_1 + \sqrt{\epsilon_1^2 - \lambda^2}} & \text{if } \epsilon_1 > \lambda
        \end{cases}.
        \label{gamma_A}
\end{align}
\mk{Similarly, $\gamma^B$ can be written as ~\cite{PhysRevB.103.174205}}
\begin{equation}
  \gamma^B(E)=\lim_{L \rightarrow \infty} \frac{1}{2\pi L} \int_0^{2\pi} \ln \left\lVert \prod\limits_{k=1}^LB_k(\theta) \right\rVert d\theta.
\end{equation}

\mk{Eq.~\eqref{Lattice_equation} is invariant under $E \to -E$, $s_k \to (-1)^k s_k$, reflecting the chiral symmetry of the model, so the spectrum is symmetric with respect to $E = 0$. The point $E = 0$ is special: Eq.~\eqref{Lattice_equation} reduces to
\begin{equation}
    \zeta_k s_{k+1} = -\zeta_{k-1}s_{k-1},
\end{equation}
so the even and odd sublattices decouple. Correspondingly, $B_k$ becomes purely off-diagonal and
\begin{equation}
    B_{k+1}B_k
    =
    \operatorname{diag}
    \left(
        -4\zeta_k^2,
        -4\zeta_{k+1}\zeta_{k-1}
    \right),
\end{equation}
which gives exactly $\gamma^B(0) = -\gamma^A$ and hence $\gamma(0) = \gamma^A+\gamma^B(0)=0$. The $E = 0$ states are therefore always delocalized: ergodic for $\epsilon_1 > \lambda$ and multifractal for $\epsilon_1 < \lambda$, where they reduce to a limiting case of the off-diagonal model of Ref.~\cite{PhysRevB.91.014108}.}

\mk{For $E\neq 0$, w}e use Avila's global theory~\cite{Avila2015}.
According to Refs.~\cite{Cai_2023, Avila2017, Jitomirskaya1994}, the presence ($\epsilon_1<\lambda$) [absence ($\epsilon_1>\lambda$)] of a singularity in the transfer matrix, Eq.~\eqref{transfer_matrix}, excludes (allows) the existence of the ergodic states in the spectrum~\footnote{Though in the case of our $E$-dependent effective Hamiltonian this result might not be mathematically rigorous.}. Thus, only AMEs are realized for $\epsilon_1<\lambda$, while only the conventional MEs appear for $\epsilon_1>\lambda$.
\imk{
}

Following Avila's approach, we analyze the large-$\epsilon$ limit of the complexified cocycle. We therefore complexify the phase, i.e. $\theta \to \theta + i\epsilon$, and take $\epsilon\to\infty$. Then $\zeta_k$ becomes
$$\zeta_k\approx\frac{\lambda}{2}e^{\epsilon}e^{-i\phi}$$
\imk{and at} \mk{large $\epsilon$, all $B_k$ share the same $k$-independent leading rank-one structure,
\[
B_k \simeq \frac{E\lambda^2}{4}e^{2\epsilon}e^{-2i\phi_k}\,P,
\qquad
P=\begin{pmatrix} 1 & 0 \\ 0 & 0\end{pmatrix},
\]
\imk{with $\phi_k=2\pi k\beta+\theta$ and the} corrections of relative order $e^{-\epsilon}$. Since the leading term is proportional to the
same \imk{
projector
matrix $P = P^2$} for all $k$, with a $k$-independent eigenvector $(1,0)^T$ and a
vanishing second eigenvalue, the product of the leading terms \imk{
reads as,}
\[
\prod_{k=1}^{L}\frac{E\lambda^2}{4}e^{2\epsilon}e^{-2i\phi_k}P
=\left(\frac{E\lambda^2}{4}\right)^{L} e^{2L\epsilon}\,
e^{-2i\sum_k \phi_k}\,P ,
\]
where the \imk{phase term has} unit modulus. Its norm is $\lVert B_k\rVert^{L}$ with
$\lVert B_k\rVert = e^{2\epsilon}\frac{|E|\lambda^{2}}{4}+O(e^{\epsilon})$, which is independent
of both $k$ and $\theta$, so that for $E \neq 0$,
\begin{equation}
 \gamma_\epsilon^{B}(E) = 2\epsilon + \ln\left|\frac{E\lambda^2}{4}\right| + O(e^{-\epsilon}).
\label{gamma_B_at_inf}
\end{equation}
This is the expression of $\gamma^B_\epsilon$ as $\epsilon\to\infty$. Avila's global theory of one
frequency analytical $SL(2,\mathbb{R})$ cocycle~\cite{Avila2015} shows that
$\gamma^B_\epsilon$ is a convex, piecewise linear function of $\epsilon$
with slope defined as:
\begin{equation}
\omega(E) =\lim_{\epsilon \to 0^+} \frac{\gamma^B_\epsilon(E) - \gamma^B_0(E)}{\epsilon},
\end{equation}
being integer-valued quantized. As we can see
from Eq.~\eqref{gamma_B_at_inf}, the slope $\omega(E)$ of $\gamma^B_\epsilon(E)$
with respect to $\epsilon$, is $2$ as $\epsilon\to\infty$. Moreover, in our case $\det B_k \neq 1$ after the complexification, i.e., $B_k$ is not in the $SL(2,\mathbb{R})$ group and therefore, according to~\cite{Jitomirskaya2013, Jitomirskaya2017}, the above quantization of $\omega(E)$ is half-integer. Since $\omega(E)$ is half-integer quantized, as
we start decreasing $\epsilon$ from $\infty$, the slope in the
neighborhood of $\epsilon=0$ can be $2$, $3/2$, $1$, $1/2$ or $0$. If there were just two limiting slopes available (like in~\cite{PhysRevB.104.134202}),
$\omega(E) = 2$ and $0$, we would have the following LE, for the case where $E$
belongs to the spectrum:
\begin{equation}
  \gamma^B_0= \ln\left|\frac{E\lambda^2}{4}\right|.
\end{equation}
Since $A_k$ is a scalar and $\gamma^{A}$ is independent of $\epsilon$, the physical Lyapunov exponent is $\gamma(E) = \gamma^{A} + \gamma^{B}_{0}$~\cite{Cai_2023}, while the acceleration $\omega(E)$ is determined entirely by $\gamma^{B}_{\epsilon}$. Moreover, by
Avila's global theory, the energy does not belong to the spectrum, iff
$\gamma(E)>0$ and $\gamma^B_\epsilon(E)$ is flat in the neighborhood
of $\epsilon=0$, i.e., $\omega(E)=0$. Following Avila's global theory \imk{(with the above reservation about the slope $2$)}, the spectrum can then be decomposed as in Table~\ref{classification}.
\begin{table}
\centering
\resizebox{0.3\textwidth}{!}{%
\begin{tabular}{|c|c|c|}
\hline
$\gamma (E)$ & $\omega_{\epsilon=0}$ & State behavior \\
\hline
$> 0$ & $= 0$ & $E \notin \Sigma$ \\
$> 0$ & $> 0$ & Localized  \\
$= 0$ & $> 0$ & Multifractal  \\
$= 0$ & $= 0$ & Ergodic \\
\hline
\end{tabular}
}
\caption{Classification of states based on $\gamma (E)$ and $\omega$.}
\label{classification}
\end{table}
The total LE, using $\gamma (E)\geq0$, can be written as
\begin{align}
  \gamma (E) &= \begin{cases}
    \max\left[\ln\left|\frac{E\lambda}{4}\right|,0\right], & \epsilon_1 < \lambda \quad \\
    \max\left[\ln\left|\frac{E \lambda^2}{4\epsilon_1 + 4\sqrt{\epsilon_1^2 - \lambda^2}}\right|,0\right]
    & \epsilon_1 > \lambda
  \end{cases}.
  \label{Final_gamma}
\end{align}}
Then the mobility edge can be determined by $\gamma(E)=0$, which gives
\begin{align}
  |E| = \begin{cases}
            \frac{4}{\lambda} & \text{if } \epsilon_1 < \lambda \quad \\
            \frac{4(\epsilon_1 + \sqrt{\epsilon_1^2 - \lambda^2})}{\lambda^2}& \text{if } \epsilon_1 > \lambda
          \end{cases}.
          \label{Final_ME}
\end{align}
Unfortunately, we have multiple additional slopes $\omega(E)$ in between, thus, with the above result we can distinguish only the localized states with slope $2$ from the rest of the states which may have localized or delocalized character.
Thus, Eq.~\eqref{Final_gamma} gives us only the lower bound for the Lyapunov exponent (as some of the other possible
  values of the slope are not included in the calculation). The expression for $\epsilon_1 = 0$ exactly matches the form of the MEs previously obtained for the ABF diamond chain with only the AA potential~\cite{aamna}, which was derived using an analogy with the extended Harper model~\cite{Avila2017}.
However, for $\epsilon_1 > 0$, our analytical expression does not fully agree with the numerical results. In addition, for $\epsilon_1<\lambda$, we will show below that our model exhibits multiple AMEs that cannot be predicted by any of the above theories. This emergence may be attributed to the massive degeneracy inherent to the ABF diamond chain, where for $\epsilon_1 < \lambda$ the energy levels remain strongly clustered, potentially modifying the spectral structure.

To the best of our knowledge, our work is the first to report the emergence of such a slope-2 behavior in this context. We find that the analytical estimate obtained from the Avila inspired approach does not provide a complete description of the mobility edge structure of the diamond chain lattice with antisymmetric disorder and constant offset. Nonetheless, we find that the MEs obtained by fitting Eq.~\eqref{Final_ME_fitting} for $\epsilon_1 > \lambda$ remain consistent with the numerical data in the large $\lambda$ limit. We will explain this fitting protocol later in this paper. Moreover, as we will see below, the overall picture of the AMEs for $\epsilon_1<\lambda$, transforming into the conventional one for $\epsilon_1>\lambda$ is consistent with the singular/regular properties of the transfer matrix, mentioned in~\cite{singularity_remove, Avila2017, Jitomirskaya1994}.

\begin{figure*}
    \centering
    \includegraphics[width=\textwidth]{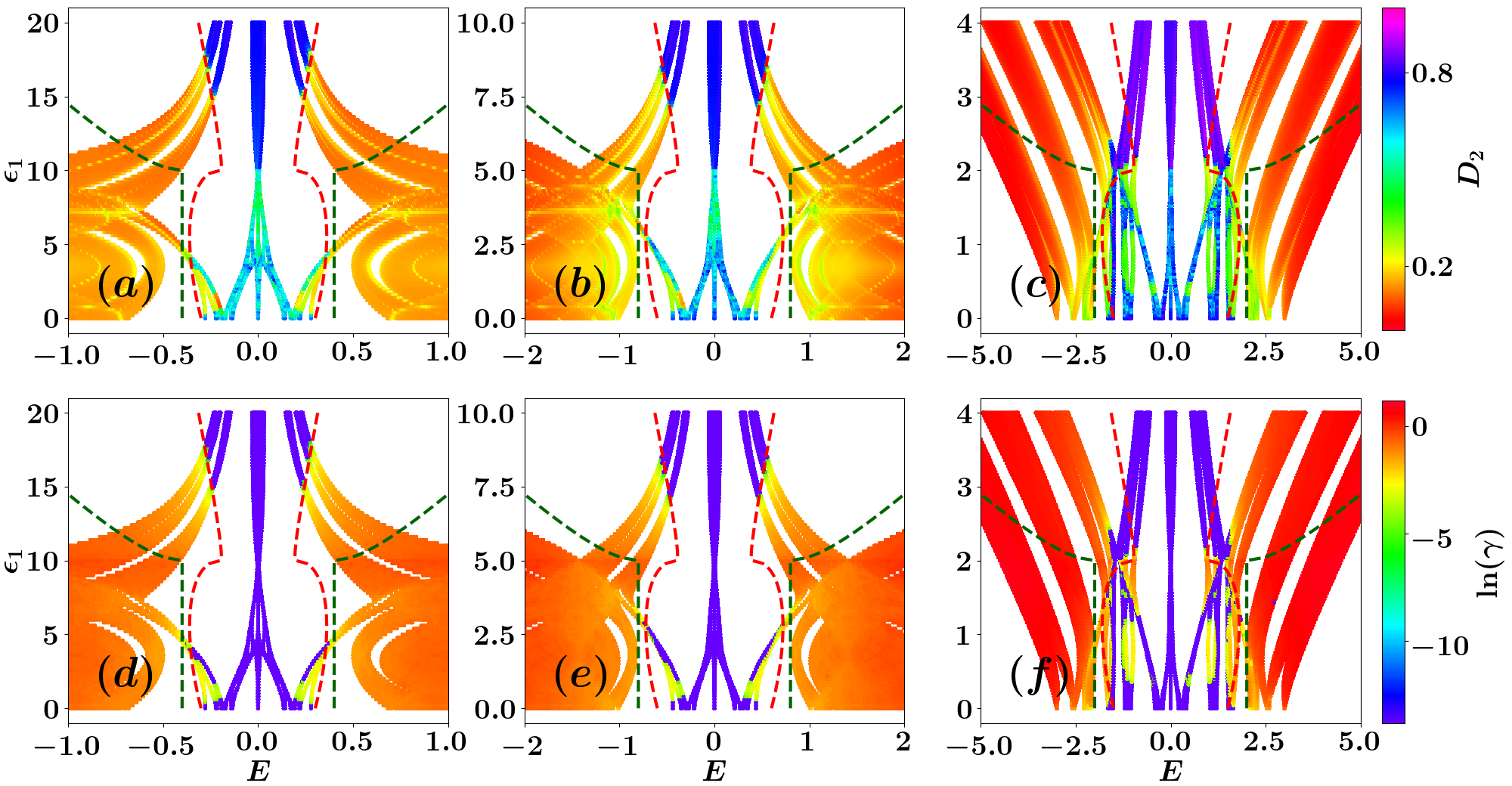}
    \caption{(a)-(c) The fractal dimension $D_2$ (whose magnitude is captured as shown in the colorbar) as a function of increasing $\epsilon_1$ and energy $E$ for (a) $\lambda=10$, (b) $\lambda=5$, and (c) $\lambda=2$. (d)-(f) The Lyapunov exponent $\gamma$ (whose magnitude in the logscale is captured as shown in the colorbar) as a function of increasing $\epsilon_1$ and energy $E$ for (d) $\lambda=10$, (e) $\lambda=5$, and (f) $\lambda=2$. Here averaging is done over $50$ values of $\theta$. The system size is $N=12543$ and $J=1$. Green dashed lines are from Avila-inspired analysis given by Eq.~\eqref{Final_ME} and red dashed lines correspond to the modified analytical expression using fitting, given by Eq.~\eqref{Final_ME_fitting}.}
    \label{lam_10_2_colour}
\end{figure*}

\section{Fractal dimension analysis}\label{sec:D2}

Next, to confirm our results, we perform a numerical analysis based on the fractal characteristics of the eigenstates. In this study, we focus on the regime with $\epsilon_1 > 0$. A standard tool for probing the localization properties of wave functions is the inverse participation ratio ($I_2$), which quantifies the extent of localization. The IPR for a given eigenstate is defined as
\begin{equation}
  I_q=\sum_{n=1}^{L}\sum_{\alpha=u,c,d}|\alpha_n(E)|^{2q}
  \label{IPR}
\end{equation}
where $\alpha_n(E)$ denotes the probability amplitude of the eigenstate with energy $E$ at the $\alpha=u,c,d$ site of the $n$-th unit cell, i.e., $\alpha_n(E) \in \{u_n(E), c_n(E), d_n(E)\}$ from Eq.~\eqref{eq:ham}. The corresponding basis states $\ket{\alpha_n}$ represent single particle states localized at site $\alpha$ in the $n$-th unit cell. IPR scales with system size $N$ as $N^{-1}(N^0)$ for ergodic (localized) states. A more complete study of localization can be done using fractal dimension $D_q$. In our analysis we focus on fractal dimension $D_2$, corresponding to $q=2$, which is given by $D_2 = -\lim_{N \to \infty} \frac{\ln(I_2)}{\ln N}$. It is known that $D_2\to 1$ for ergodic states, $D_2 \to 0$ for localized states and $D_2$ lies between 0 and 1 for multifractal states. Fig.~\ref{lam_10_2_colour}(a)-(c) shows the energy-resolved $\theta$-averaged $D_2$ as a function of shift in potential $\epsilon_1$ for the entire spectrum for potential strength $\lambda=10$, $\lambda=5$, and $\lambda=2$, respectively. Green dashed lines are given by Eq.~\eqref{Final_ME} which should separate the localized states with the slope $\omega(E)=2$ from the remaining states. Numerically, the ergodic-localized (MEs) and multifractal-localized anomalous mobility edges (AMEs) can be observed for $\epsilon_1 > \lambda$ and $\epsilon_1 < \lambda$, respectively in full agreement with~\cite{singularity_remove, Avila2017, Jitomirskaya1994}. The emergence of multifractal states in the regime $\epsilon_1 < \lambda$ is due to singularities in the transfer matrix $T_k$, which arise when the quasiperiodic coefficient $\zeta_k$ vanishes. Since $\zeta_k$ appears in the denominator of the matrix elements, its zeros lead to singular behavior of the transfer matrix at specific lattice sites.\footnote{This mechanism of small (evenly distributed) zeros in the hopping terms, present, e.g., in uncorrelated Anderson models on hierarchical graphs, also leads to the emergence of multifractality, see, e.g.~\cite{PhysRevB.110.174202}.} In the thermodynamic limit, these zeros are distributed in an incommensurate manner, affecting the spatial structure of the eigenstates. As a result, the system cannot support ergodic states in this regime. Instead, depending on the Lyapunov exponent, the spectrum consists of either localized states with finite Lyapunov exponent or critical (multifractal) states characterized by vanishing Lyapunov exponent. Our results confirm the presence of AMEs in the $\epsilon_1 < \lambda$ regime. In contrast, for $\epsilon_1 > \lambda$, $\zeta_k$ does not vanish and the transfer matrix remains regular hence the ergodic states are allowed and we observe conventional MEs separating localized from ergodic states.

Another noteworthy observation is the discrepancy between the numerical ME structure and the analytical estimate given by Eq.~\eqref{Final_ME}, as shown in Fig.~\ref{lam_10_2_colour}(a)-(c). The green dashed lines correspond to the mobility edges estimated from the Avila-inspired analysis. In the $\epsilon_1>\lambda$ regime, reddish regions corresponding to localized states appear between the estimated mobility edges. Similarly, for $\epsilon_1<\lambda$, the yellowish and greenish regions indicate localized states between the estimated AMEs. For $\lambda=0.8$, as shown in Appendix~\ref{app:lambda=0.8}, this discrepancy becomes even more pronounced. We note that the present energy-dependent quasiperiodic model lies outside the standard setting in which Avila's theory has been rigorously established. These results indicate the presence of localized states associated with acceleration slopes $\omega(E)$ different from $2$, which are not captured by Eq.~\eqref{Final_ME}. We confirm this conclusion through Lyapunov exponent analysis in the following section.

\section{Lyapunov Exponent analysis}\label{sec:Lyapunov Exponent}

To further validate the results obtained from the fractal dimension analysis, we numerically examine the Lyapunov exponent (LE). The LE, denoted by $\gamma(E)$, quantifies the exponential rate at which a wave function localizes around the location of its maximum $n_0$. For a localized state, the probability density decays as
\begin{equation}
|\psi_n(E)|^2 \sim \exp\left[-2\gamma(E)\, |n - n_0|\right],
\label{Expo_decay_main}
\end{equation}
where a finite $\gamma(E) > 0$ indicates exponential localization. In contrast, $\gamma(E) = 0$ indicates the absence of exponential localization and may correspond to either ergodic or multifractal states. We compute $\gamma(E)$ for all eigenstates by fitting their spatial wave function profiles using Eq.~\eqref{Expo_decay_main}. Numerically, both ergodic and multifractal states yield $\gamma \approx 0$.

\begin{figure}
    \centering
    \includegraphics[width=0.48\textwidth]{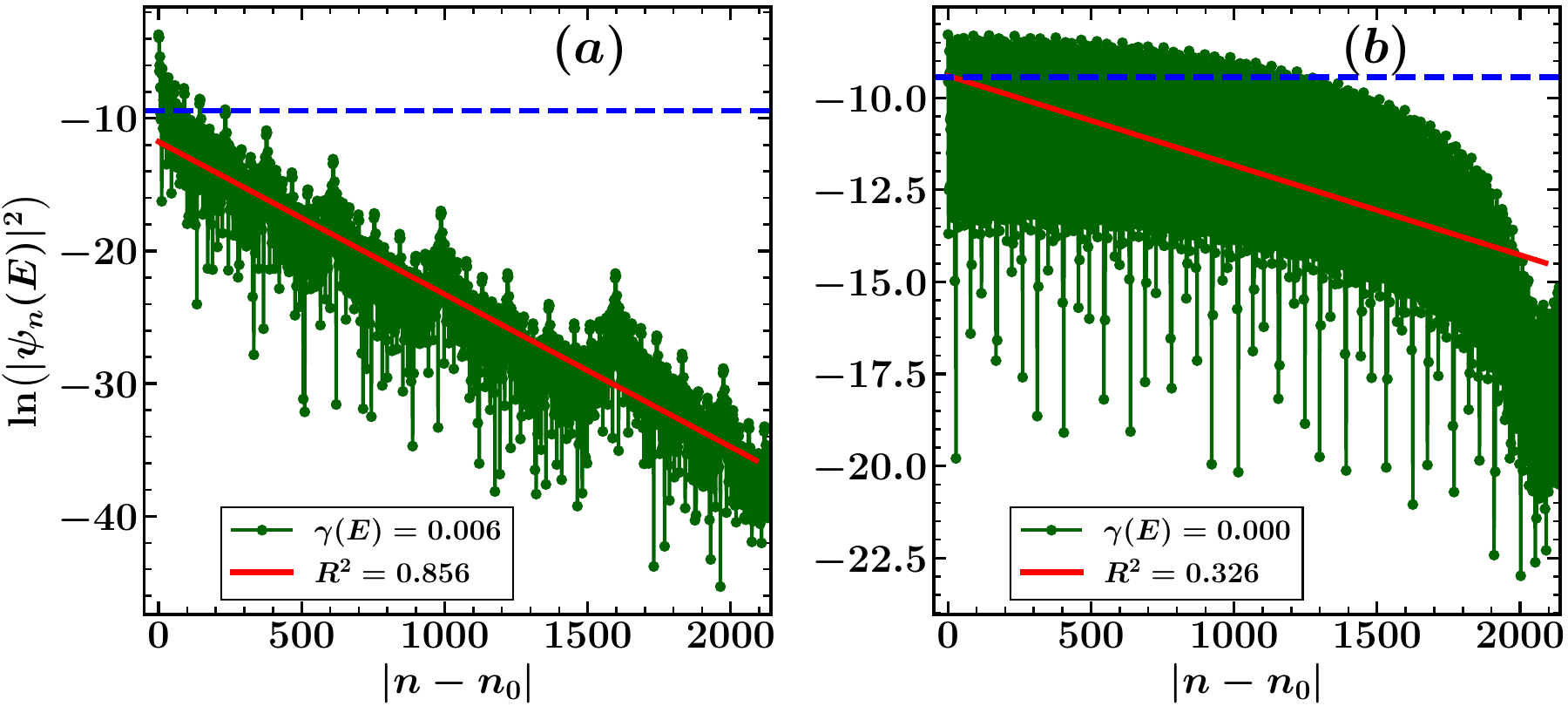}
\caption{
Spatial decay profiles of two consecutive eigenstates.
(a) A localized state with $E=-0.26$ exhibiting clear exponential decay, with $\gamma(E)=0.006$ and a good linear fit ($R^2=0.856$).
(b) The neighboring delocalized state with $E=-0.22$, characterized by $\gamma(E)\approx 0$ and poor linear fitting quality ($R^2=0.326$), indicating the absence of exponential localization.
The dashed horizontal line marks the reference level $-\ln N$ of the ergodic wave function. Here averaging is done over 50 values of $\theta$. Number of unit cells $L=4181$, system size $N=3L=12543$, $\lambda=10$, and $\epsilon_1=16.8$.
}

    \label{Rc_figure}
\end{figure}

To ensure the robustness of fitting, we evaluate the goodness of fit parameter $R^2$ for each eigenstate. For localized states the fits are robust with $R^2\approx 1$. However in the multifractal and ergodic regimes, the wavefunctions do not show exponential decay, resulting in significantly reduced values of $R^2$. We therefore use the quality of the fit itself as a diagnostic to separate localized from delocalized states. We choose the critical value of $R_c^2=0.85$; the states with $R^2\geq R_c^2$ are classified as exponentially localized and the states with $R^2<R_c^2$ are identified as delocalized. For the latter, the extracted $\gamma$ has no physical meaning since the exponential form is not applicable, hence we set $\gamma$ to a negligibly small value ($10^{-14}$). To justify the chosen threshold $R_c^2$, we examine the wavefunction profiles of two consecutive eigenstates across the localization transition. For the localized state [Fig.~\ref{Rc_figure}(a)] we obtain $R^2 = 0.856$, whereas for the subsequent delocalized state [Fig.~\ref{Rc_figure}(b)] the goodness-of-fit parameter drops sharply to $0.326$. This procedure ensures that only states exhibiting genuine exponential localization contribute finite Lyapunov exponents.

Figures~\ref{lam_10_2_colour}(d)–(f) show the $\theta$-averaged Lyapunov exponent $\gamma$ as a function of $\epsilon_1$ and energy $E$ for $\lambda = 10$, $5$, and $2$, respectively. These phase plots exhibit the same qualitative features observed in the fractal dimension figures. For $\lambda = 10$ and $5$, in the regime $\epsilon_1 < \lambda$, we observe greenish regions between the analytically obtained MEs, where $\ln(\gamma)$ becomes slightly negative, indicating the presence of localized states. A similar behavior is observed for $\lambda = 2$, where localized states emerge between the green dashed lines. Although the Lyapunov exponent alone cannot distinguish between ergodic and multifractal states, since $\gamma(E) = 0$ in both cases, a combined analysis with the phase diagrams of the fractal dimension $D_2$ [Figs.~\ref{lam_10_2_colour}(a)–(c)] confirms the presence of multifractal states in the regime $\epsilon_1 < \lambda$. To further confirm the presence of localized states as well as multifractal states between our predicted MEs in the $\epsilon_1<\lambda$ regime, we further analyze their individual spatial decay profiles.

\begin{figure*}
    \centering
    \includegraphics[width=\textwidth]{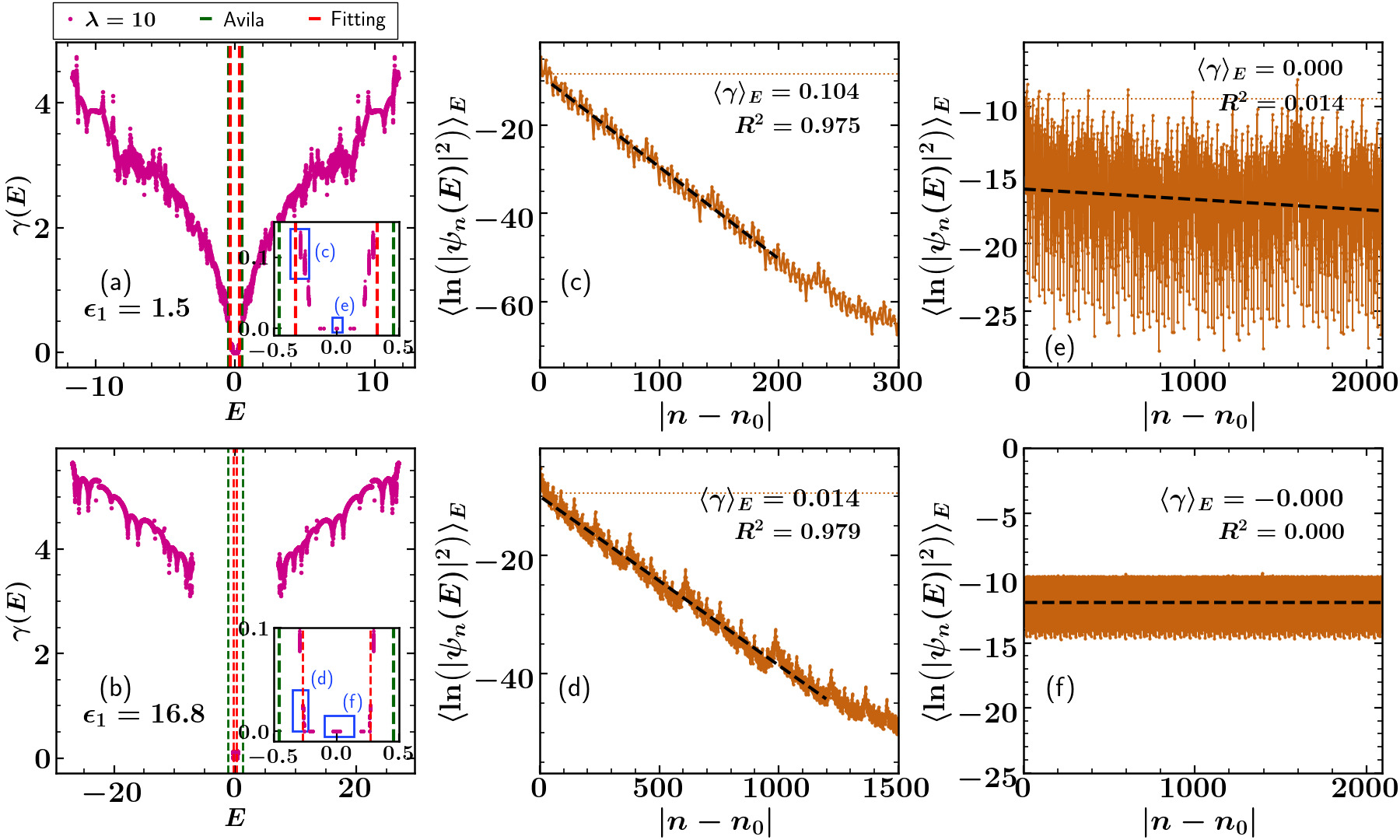}

\caption{
\mk{(a), (b) Lyapunov exponent $\gamma(E)$ across all eigenstates for (a) $\epsilon_1 = 1.5$ and (b) $\epsilon_1 = 16.8$. Green dashed lines mark the analytical estimate of the mobility edges motivated by Avila's global theory [Eq.~\eqref{Final_ME}], while red dashed lines are plotted using Eq.~\eqref{Final_ME_fitting} obtained after the fitting procedure.
(c)–(f) Averaged $\ln(|\psi_n(E)|^2)$ over eigenstates with energies within the specified intervals, plotted as a function of distance $|n - n_0|$ from the site of maximum amplitude $n_0$.
(c),~(d)~Profiles averaged over (c) $E \in [-0.3, -0.25]$ ($\epsilon_1 = 1.5$) and (d) $E \in [-0.28, -0.25]$ ($\epsilon_1 = 16.8$). Black dashed lines show exponential fits [Eq. \eqref{Expo_decay_main}] yielding $\langle \gamma\rangle_E = 0.104$ ($R^2 = 0.975$) and $0.014$ ($R^2 = 0.979$), respectively, confirming robust exponential decay and localized states.
(e),~(f)~Profiles averaged over (e) $E \in [-0.004, 0.004]$ ($\epsilon_1 = 1.5$) and (f) $E \in [-0.06, 0.06]$ ($\epsilon_1 = 16.8$). Exponential fits yield $\langle \gamma\rangle_E \approx 0$ ($R^2 \approx 0$) in both cases, signifying delocalized behavior.
Blue rectangles in panels (a) and (b) highlight the corresponding energy windows. Averages are taken over 50 realizations of phase $\theta$, with $N = 12543$, $\lambda = 10$, and $J = 1$.}
}

    \label{spatial_decay_lam_10}
\end{figure*}

\subsection{Spatial decay profiles}

We begin by examining the regime $\epsilon_1 < \lambda$, where singularities appear in the transfer matrix of the effective one-dimensional model. To validate the observations from the phase diagrams, we plot the Lyapunov exponent $\gamma(E)$ for $\lambda = 10$ and $\epsilon_1 = 1.5$ [Fig.~\ref{spatial_decay_lam_10}(a)]. The green dashed lines denote the AMEs given by Eq.~\eqref{Final_ME}. Nonzero values of $\gamma(E)$ are observed between these green lines, indicating localized behavior of the eigenstates and confirming the inability of Avila's theory to locate the ME. In the $\epsilon_1 > \lambda$ regime, transfer matrix singularities are absent, and we observe conventional MEs separating localized and ergodic states [Fig.~\ref{spatial_decay_lam_10}(b)]. Even in this regime, small but significantly non-zero $\gamma(E)$ values appear between the green lines of Eq.\eqref{Final_ME}, demonstrating a quantitative deviation from the numerical results. To further confirm the nature of the eigenstates, we analyze the spatial decay of the wavefunction amplitudes by averaging $\langle \ln(|\psi_n(E)|^2)\rangle_E$ over eigenstates within specific energy intervals for various values of $\lambda$, where $\langle \cdot \rangle_E$ denotes averaging over eigenstates within the chosen energy interval.

\mk{For localized states, the averaged profile $\langle \ln|\psi_n(E)|^2 \rangle_E$, obtained by aligning the maxima of the individual eigenstates at $n_0$ and averaging over eigenstates within the chosen energy window and over realizations of $\theta$, decays linearly in $|n-n_0|$, in agreement with the exponential form of Eq.~\eqref{Expo_decay_main}. For delocalized (multifractal or ergodic) states, in contrast, no such linear decay is observed: the averaged profile either remains flat or shows strong spatial fluctuations and revivals, so that an exponential fit fails and yields a low goodness of fit $R^2$. Such states are therefore assigned $\gamma(E) \approx 0$. These averaged profiles serve only as a qualitative illustration of the presence or absence of exponential localization. The distinction between multifractal and ergodic states is based on the fractal dimension $D_2$ [Figs.~\ref{lam_10_2_colour}(a)-(c)] and on the finite-size scaling of $\langle\gamma\rangle_E$ and $I_2$ (Appendices~\ref{app:Lyapunov_exp_vs_N} and \ref{app:IPR_collapse}).}

Figs.~\ref{spatial_decay_lam_10}(c) and ~\ref{spatial_decay_lam_10}(d)
represent this average for localized eigenstates with energies in the
range $E \in [-0.3, -0.25]$ at $\epsilon_1 = 1.5$, and energies in the
range $E \in [-0.28, -0.25]$ at $\epsilon_1 = 16.8$, shown by blue
rectangles in the inset in panels (a) and (b), for $\lambda=10$ with
$\langle \gamma\rangle_E=0.104$ ($R^2 = 0.975$), and $\langle
\gamma\rangle_E=0.014$ ($R^2 = 0.979$), respectively, exhibiting clear
exponential decay. The corresponding positive LE, significantly larger
than the inverse system size $1/N$, confirms the localized nature of
these states. \mk{In contrast, Fig.~\ref{spatial_decay_lam_10}(e)
  shows the same analysis for eigenstates within the energy range
  $[-0.004, 0.004]$ at $\epsilon_1 = 1.5$, for $\lambda = 10$. Here
  the averaged profile shows no systematic decay but instead shows
  strong spatial fluctuations, so that the exponential fit fails,
  yielding $\langle \gamma\rangle_E = 0.000$ with a poor goodness of
  fit ($R^2 = 0.014$). These states are therefore delocalized, and
  combined with the fractal dimension analysis of
  Figs.~\ref{lam_10_2_colour}(a)-(c), they are identified as
  multifractal.} The existence of localized states within the interval
between the green lines of Eq.~\eqref{Final_ME} indicates that the
analytical estimate obtained from the Avila-inspired approach does not
provide a complete description of the mobility edge structure in the
present energy dependent quasiperiodic model.  Furthermore,
Fig.~\ref{spatial_decay_lam_10}(f), for $E \in [-0.06, 0.06]$ at
$\epsilon_1 = 16.8$, shows a flat profile with $\langle
\gamma\rangle_E = 0.000$ ($R^2 = 0.000$), indicating ergodic behavior
of states in the $\epsilon_1>\lambda$ regime which is consistent with
our phase diagrams.  Hence we can conclude the presence of AMEs in the
$\epsilon_1<\lambda$ regime and MEs in the $\epsilon_1>\lambda$
regime. We further confirm these results by studying $\langle \gamma
\rangle_E$ with system size in Appendix~\ref{app:Lyapunov_exp_vs_N}
using exact diagonalization. We have also confirmed these results by
computing the LE directly using the transfer matrix, which yields
consistent results.

\section{Disagreement with Eq.~(\ref{Final_ME})}\label{sec:Avila_discrepancy}

In this section, we discuss in detail the discrepancy between the numerical mobility edge structure and the analytical estimate motivated from Avila's global theory given by Eq.~\eqref{Final_ME}. Our findings indicate that the analytical estimate agrees only qualitatively but fails quantitatively. To achieve quantitative agreement, a fitting parameter must be introduced into the expression motivated by Avila's global theory. The following subsection outlines the fitting protocol used for this purpose.

\begin{figure}
    \centering
    \includegraphics[width=0.48\textwidth]{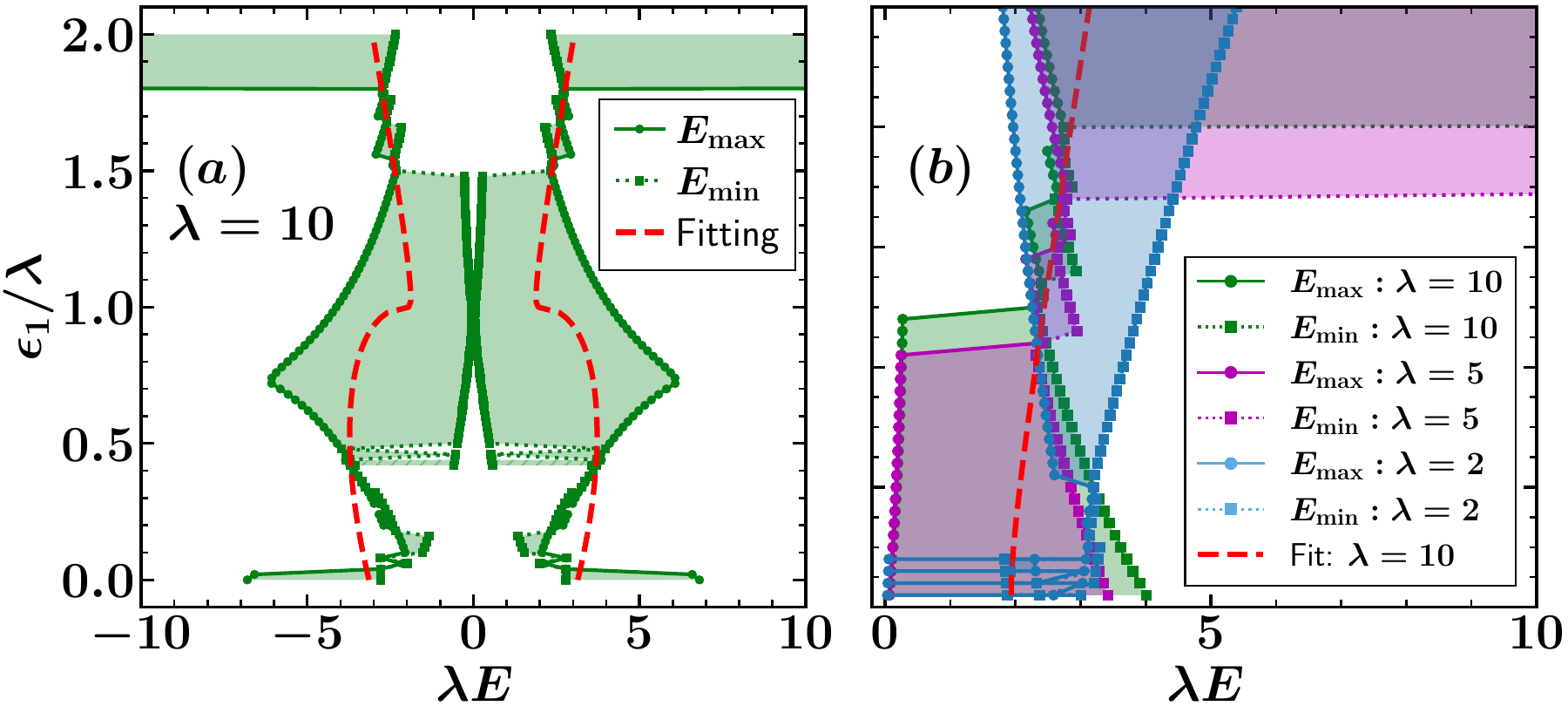}
\caption{(a) Offset $\epsilon_1$ versus eigenenergy $E$ showing the numerically obtained
mobility edges $E_{\text{min}}$ and $E_{\text{max}}$ for $\lambda=10$.
(b) Comparison of $E_{\text{min}}$ and $E_{\text{max}}$ in $\epsilon_1>\lambda$ regime for $\lambda=10$, $5$, and $2$,
demonstrating the scaling form of the fitting expression in Eq.~\eqref{Final_ME_fitting}. Red dashed lines denote the fitting curves. The colored shaded region between
$E_{\text{min}}$ and $E_{\text{max}}$ marks the allowed region for the mobility edge,
with the shade color same as the corresponding $\lambda$ curves. All quantities are averaged over $50$ realizations of the phase parameter $\theta$. The system size is $N=12543$ and $J=1$.}
    \label{ep1_vs_E_fitting}
\end{figure}

\begin{figure*}
    \centering
    \includegraphics[width=\textwidth]{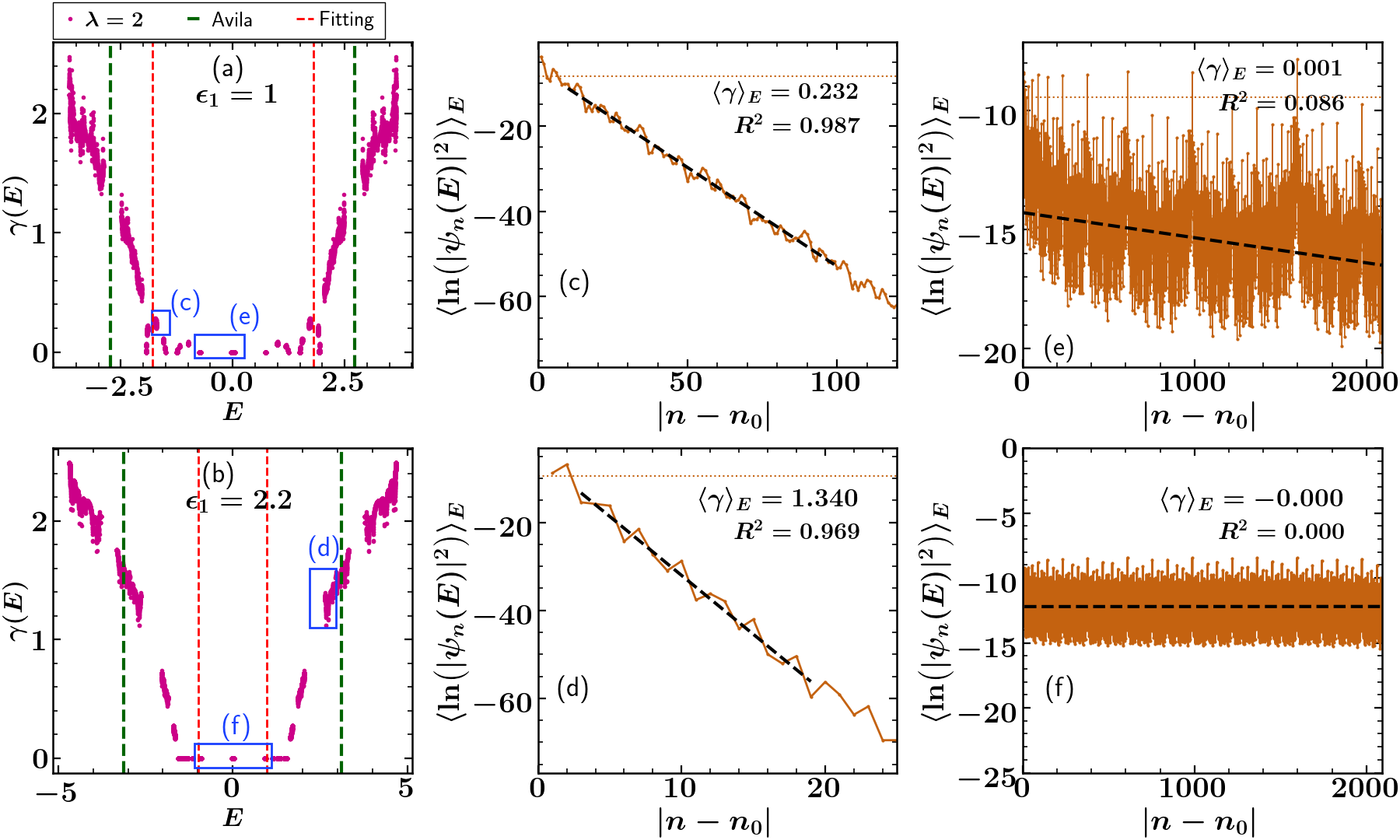}

\caption{
\mk{(a), (b) Lyapunov exponent $\gamma(E)$ across all eigenstates for (a) $\epsilon_1 = 1$ and (b) $\epsilon_1 = 2.2$. Green dashed lines mark the analytical estimate of the mobility edges motivated by Avila's global theory [Eq.~\eqref{Final_ME}], while red dashed lines are plotted using Eq.~\eqref{Final_ME_fitting} obtained after the fitting procedure.
(c)–(f) Averaged $\ln(|\psi_n(E)|^2)$ over eigenstates with energies within the specified intervals, plotted as a function of distance $|n - n_0|$ from the site of maximum amplitude $n_0$.
(c),~(d)~Profiles averaged over (c) $E \in [-1.8, -1.6]$ ($\epsilon_1 = 1$) and (d) $E \in [2.5, 3.0]$ ($\epsilon_1 = 2.2$). Black dashed lines show exponential fits [Eq. \eqref{Expo_decay_main}] yielding $\langle \gamma\rangle_E = 0.232$ ($R^2 = 0.987$) and $1.340$ ($R^2 = 0.969$), respectively, confirming robust exponential decay and localized states.
(e),~(f)~Profiles averaged over (e) $E \in [-0.75,0.25]$ ($\epsilon_1 = 1$) and (f) $E \in [-1.0, 1.0]$ ($\epsilon_1 = 2.2$). Exponential fits yield $\langle \gamma\rangle_E \approx 0$ ($R^2 \approx 0$) in both cases, signifying delocalized behavior.
Blue rectangles in panels (a) and (b) highlight the corresponding energy windows. Averages are taken over 50 realizations of phase $\theta$, with $N = 12543$, $\lambda = 2$, and $J = 1$.}
}

    \label{spatial_decay_lam_2}
\end{figure*}

\subsection{Quantitative disagreement: Fitting Protocol}
To obtain an approximate \mk{analytical estimate of the mobility edge, motivated by Avila’s global theory} by a fitting form similar to Eq.~\eqref{Final_ME}, we analyze the phase plots of the Lyapunov exponent $\gamma(E)$, extracted from exponential fits to the spatial decay of individual eigenstates. As discussed above, only states exhibiting a robust exponential fit are assigned a finite Lyapunov exponent, while delocalized states are identified based on a low goodness of fit parameter $R^2$. Using this classification, we determine the mobility edges as follows. For each value of $\epsilon_1$, we identify the highest absolute value of energy belonging to the delocalized regime and denote it by $E_{\text{min}}$, and the lowest absolute value of energy belonging to the localized regime, denoted by $E_{\text{max}}$. In Fig.~\ref{ep1_vs_E_fitting}(a) we plot both $E_{\text{min}}$ and $E_{\text{max}}$ as functions of $\epsilon_1$; the points where these two curves meet define the approximate locations of the mobility edges. These extracted points are then fitted using a functional form inspired by Eq.~\eqref{Final_ME}, which captures the qualitative dependence of the mobility edge on $\epsilon_1$ and $\lambda$. The fitting is performed in the $\epsilon_1 > \lambda$ regime, where singularities are absent and the numerical behavior is more consistent with the assumptions of Avila's theory analysis. For $\lambda = 10$, in the $\epsilon_1 > \lambda$ regime we fit our data to the functional form
\begin{equation}
E = \frac{X \epsilon_1 + Y \sqrt{|\epsilon_1^2 - \lambda^2|}}{\lambda^2}.
\label{fit_form}
\end{equation}
The best-fit parameters are $X = 1.92 \pm 0.15$ and $Y = -0.46 \pm 0.15$ with goodness of fit parameter $R^2=0.98$. Since the mobility edge must remain continuous across the boundary at $\epsilon_1 = \lambda$, we impose the condition
\begin{align*}
E(\epsilon_1 = \lambda + 0) &= E(\epsilon_1 = \lambda - 0), \\
X_1 &= X_2.
\end{align*}
Accordingly, we fix $X_2 = X_1$ and fit only the coefficient $Y_2$ in the regime $\epsilon_1 < \lambda$, obtaining $Y_2 = 2.94 \pm 0.10$ with goodness of fit parameter $R^2=0.40$. The resulting fitted expression for the mobility edges is
\begin{equation}
|E| =
\begin{cases}
\dfrac{2\epsilon_1 + 3\sqrt{\lambda^2 - \epsilon_1^2}}{\lambda^2}, & \text{if } \epsilon_1 < \lambda, \\[2ex]
\dfrac{2\epsilon_1 - 0.5\sqrt{\epsilon_1^2 - \lambda^2}}{\lambda^2}, & \text{if } \epsilon_1 > \lambda.
\end{cases}
\label{Final_ME_fitting}
\end{equation}
To test the validity of this fitted form, we compare it with the numerically obtained $E_{\text{max}}$ and $E_{\text{min}}$ for $\lambda=10$, $5$, and $2$, in the $\epsilon_1>\lambda$ regime only, in the same plot in Fig.~\ref{ep1_vs_E_fitting}(b). Using the scaling form of the fitted ME in Eqs. \eqref{fit_form}-\eqref{Final_ME_fitting}, we plot $E_{\min}$ and $E_{\max}$ for several $\lambda$ values on the same plot, after the rescaling $(\lambda E, \epsilon_1/\lambda)$, collapsing the analytical results for all $\lambda$ values. We have shown data and fitting only for $E>0$, as energies are symmetric around $E=0$. The shaded region between $E_{\text{min}}$ and $E_{\text{max}}$ marks the allowed region for the mobility edge (red fitted line). For $\lambda=10$, and $5$ the fitted curve clearly separates $E_{\text{max}}$ from $E_{\text{min}}$, indicating good agreement with numerical results. In contrast, for $\lambda=2$, the fitted expression fails to provide such a separation signaling a breakdown of the fitting in this regime. The same behavior is also evident in the phase diagrams shown in Fig.~\ref{lam_10_2_colour}. In particular, for larger values of $\lambda$ ($\lambda = 10$ and $5$), the fitted mobility edges (red dashed lines) reproduce the numerical phase boundaries reasonably well in the $\epsilon_1 > \lambda$ regime. However, in the $\epsilon_1 < \lambda$ regime, the fit fails to completely separate localized states from multifractal ones
[red dashed lines in Fig.~\ref{spatial_decay_lam_10}(a)], since the functional form used for fitting is derived specifically
for the $\epsilon_1 > \lambda$ region. Moreover, for smaller values of $\lambda$ (e.g., $\lambda = 2$), the fitted expression no longer reproduces the observed ME structure and becomes unreliable.
This indicates a qualitative inapplicability of Avila's theoretical framework for $\lambda \lesssim 2$. We further examine this deviation through the spatial decay analysis for $\lambda = 2$ in the next subsection.

\subsection{Qualitative Disagreement: $\lambda=2$}

Here, we examine the qualitative disagreement of analytical estimate to locate the MEs for $\lambda \lesssim 2$.
From the phase plots for $\lambda = 2$, we observe that both the expressions, the analytical estimate of Eq.~\eqref{Final_ME} (green dashed lines)
and those obtained after fitting Eq.~\eqref{Final_ME_fitting} (red dashed lines) fail to separate localized and ergodic states in the $\epsilon_1 > \lambda$ regime
[Figs.~\ref{lam_10_2_colour}(c) and (f)].
In contrast, in the $\epsilon_1 < \lambda$ regime, not only do these analytical expressions fail to remain exact\footnote{This happens due to the breakdown of a universal form of Eq.~\eqref{fit_form} with $\lambda E$ being a function of the only parameter of $\epsilon/\lambda$, see Fig.~\ref{ep1_vs_E_fitting}(b).},
but also fail to predict the presence of multiple mobility edges. To confirm these observations, we analyze the spatial decay of eigenstates for various energy ranges for $\lambda = 2$.
Figure~\ref{spatial_decay_lam_2}(a) corresponds to $\epsilon_1 = 1$, which lies in the $\epsilon_1 < \lambda$ regime.
We observe the presence of AMEs separating localized and multifractal states.
Across the energy spectrum, the Lyapunov exponent $\gamma$ exhibits multiple points where it vanishes and reappears,
indicating the presence of several AMEs. The AME between localized  and multifractal states is further confirmed by examining the spatial decay of wave functions
at different energies [Figs.~\ref{spatial_decay_lam_2}(c) and \ref{spatial_decay_lam_2}(e)]. In contrast, for $\epsilon_1 = 2.2$ (in the $\epsilon_1 > \lambda$ regime), both localized and ergodic states are separated by the conventional ME
[Figs.~\ref{spatial_decay_lam_2}(d) and \ref{spatial_decay_lam_2}(f)].
As shown in Fig.~\ref{spatial_decay_lam_2}(b), both the analytical and fitted expressions fail to separate these states.
Collectively, these results demonstrate that for $\lambda \lesssim 2$, the analytical estimate motivated by Avila's theory exhibits a qualitative discrepancy with the numerical results and does not capture the full mobility edge structure.\footnote{Note that already at $\lambda=10$ the multiple AMEs start to emerge, see, e.g., the results at $\epsilon_1 = 4.4$ in Appendix~\ref{app:LE_4.4_multiple}.} This disagreement with Eq.~\eqref{Final_ME} is further supported by the IPR collapse analysis presented in Appendix~\ref{app:IPR_collapse}. Results for a single realization of $\theta$ are presented in Appendix~\ref{app:single_realization}, confirming that our conclusions do not depend on phase averaging.

\section{Conclusion}\label{sec:summary}

We investigate the emergence and nature of mobility edges in the ABF diamond chain subjected to a quasiperiodic Aubry-Andr\'e potential with an additional constant offset, where the potential is applied antisymmetrically to top and bottom sites.
Our study reveals two distinct types of mobility edges: (i) anomalous mobility edges separating localized and multifractal states in the regime $\epsilon_1 < \lambda$, and (ii) conventional MEs separating localized and ergodic states in the regime $\epsilon_1 > \lambda$.
These results are obtained through fractal dimension analysis and are further supported by numerical calculations of the Lyapunov exponent using both exact diagonalization and the transfer matrix method.
\mk{To further characterize the nature of the eigenstates, we analyze the spatial decay profiles of the wave-function amplitudes, which show exponential decay for localized states and the absence of such decay for multifractal and ergodic ones.}

Additionally, we derive an analytical estimate for the MEs motivated by Avila's global theory. Unfortunately, this expression is only the upper bound for the absolute value of ME energy, as it separates only the localized states with the $\epsilon$-slope $2$ from the other (localized and/or delocalized) states.
While the expression reproduces previous results for systems without an offset, the theory agrees only qualitatively for $\lambda \gtrsim 2$ and fails quantitatively when an offset is introduced.
To achieve quantitative agreement, we incorporate fitting parameters into the analytical formula, which successfully captures the numerical results for $\lambda \gtrsim 2$ in $\epsilon_1>\lambda$ regime. However, even after fitting, the analytical formula fails quantitatively in the regime $\epsilon_1 < \lambda$.
For $\lambda\lesssim2$, the numerical results reveal multiple mobility edges in the $\epsilon_1<\lambda$ regime that are not captured by the analytical estimate, leading to a qualitative discrepancy with the numerical phase diagram. This behavior may be related to the massive degeneracy inherent to the all-bands-flat diamond chain. Although the antisymmetric on-site potential lifts the exact degeneracy, for small $\epsilon_1$ and small $\lambda$ the energy levels remain strongly clustered, reflecting their flat-band origin. Our results therefore indicate that the present energy dependent quasiperiodic flat band model lies beyond the standard domain of applicability of Avila's theory and motivate the development of analytical approaches capable of describing mobility edge formation in such systems.

\section*{Acknowledgments}
We acknowledge Svetlana Jitomirskaya for the clarification of origins of
the inapplicability of the Avila's theory to our model. We are grateful to
the High Performance Computing (HPC) facility at IISER Bhopal, where
most of calculations in this project were run. M.K is grateful to the
Council of Scientific and Industrial Research (CSIR), India, for his
PhD fellowship.
\imk{I.M.K. acknowledges support by the French National Research Agency (ANR) under the France 2030 program,
in subprograms "Choose France for Science", reference ``ANR-25-CFFS-0019'', and
the TIRIS (Toulouse Initiative for Research's Impact on Society) Blockbuster Programme, reference ``ANR-22-EXES-0015'', the Occitanie region, the European Regional Development Fund (ERDF), and the European Union NextGenerationEU.}

\begin{figure}
    \centering
    \includegraphics[width=0.48\textwidth]{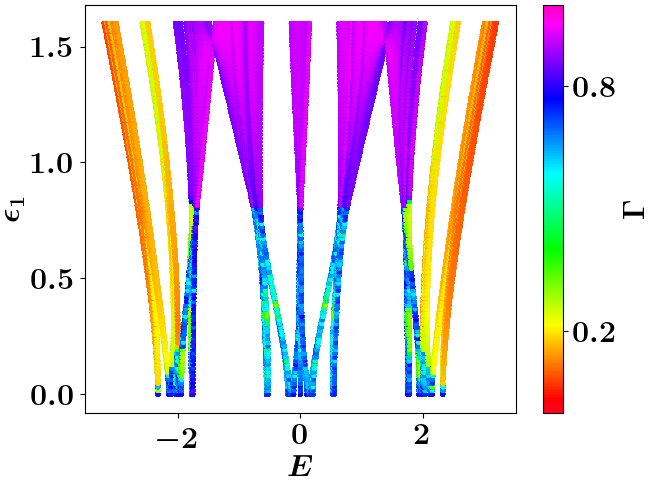}

\caption{$\epsilon_1$–$E$ phase diagram for $\lambda = 0.8$, where the color scale represents the fractal dimension.
Data is averaged over 50 realizations of the phase parameter $\theta$, with $N = 12543$ and $J = 1$. The analytical estimate, Eq.~\eqref{Final_ME}, are not visible as it lies far outside the spectral boundaries.
}

    \label{0.8_phase_plot}
\end{figure}

\appendix
\section{Phase plot for $\lambda=0.8$}\label{app:lambda=0.8}
To further illustrate the qualitative discrepancy between the numerical results and the analytical estimate of Eq.~\eqref{Final_ME} for small values of $\lambda$,
we plot the $\epsilon_1$-$E$ phase diagram with the fractal dimension $D_2$ for $\lambda = 0.8$, as shown in Fig.~\ref{0.8_phase_plot}.
We observe both multifractal and localized states in the $\epsilon_1 < \lambda$ regime, and ergodic as well as localized states in the $\epsilon_1 > \lambda$ regime.
In contrast, the analytical predictions of Eq.~\eqref{Final_ME} are not visible in the phase plot, as the corresponding mobility edge lines lie outside the spectral boundaries,
implying that all states are multifractal for $\epsilon_1<\lambda$.
A similar behavior is observed for $\epsilon_1 = 0$; for $\lambda \lesssim 1.5$, all states become multifractal.
These results highlight a significant qualitative disagreement between the numerical results and the Avila-inspired analytical estimate in the small $\lambda$ regime.

\begin{figure}
    \centering
    \includegraphics[width=0.48\textwidth]{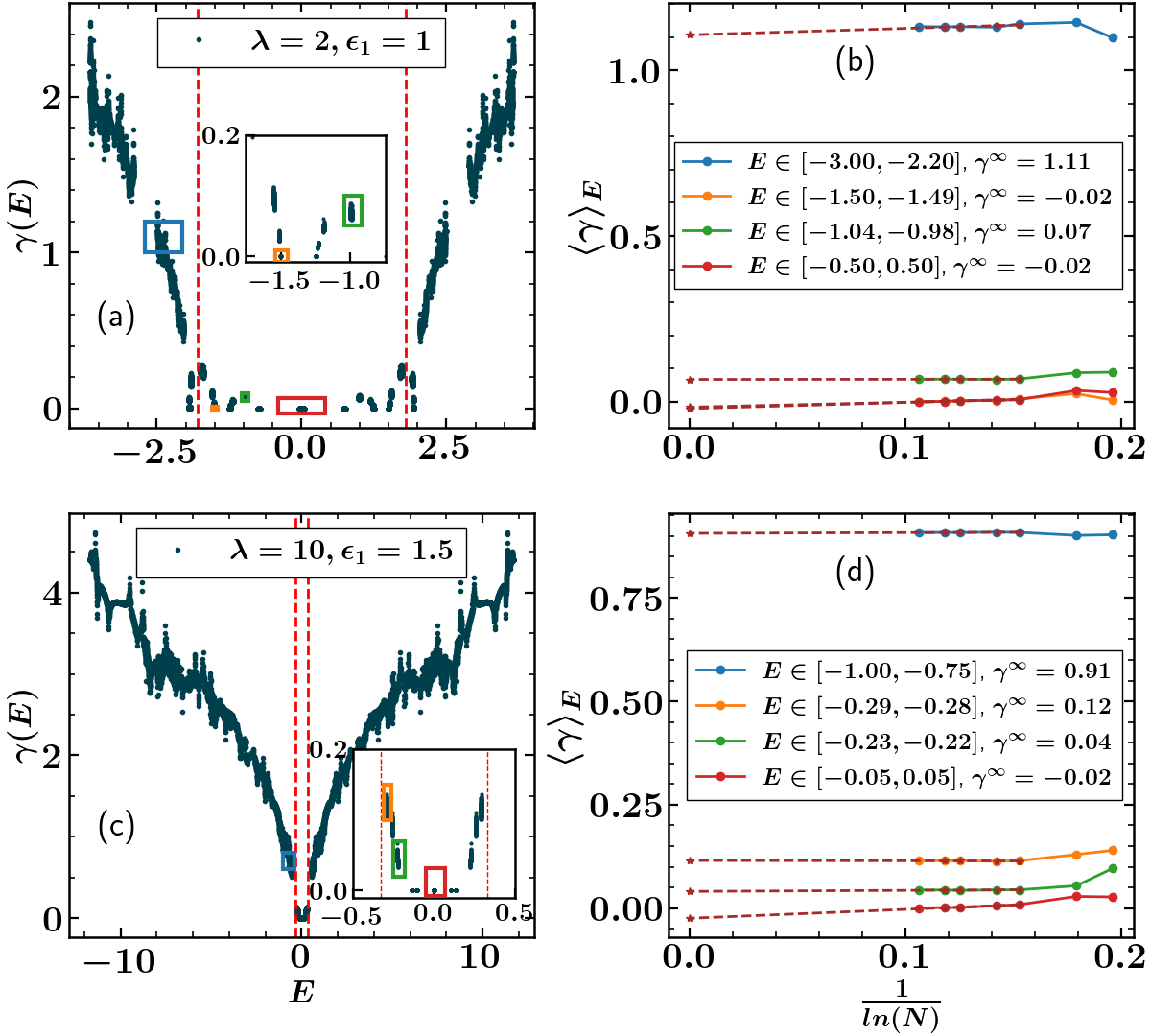}
\caption{Average Lyapunov exponent analysis with system size for $\epsilon_1<\lambda$ regime. For $\lambda=2$, (a) $\gamma(E)$ for all eigenstates with $\epsilon_1 = 1$ and (b) average $\langle \gamma \rangle_E$ with $1/\ln N $ in different energy ranges. Panels (c,d) show the same quantities for $\lambda=10$ with $\epsilon_1=1.5$. In panel (a) and (c), red dashed lines are the fitted expression Eq.~\eqref{Final_ME_fitting}. Different energy windows used in panels (b) and (d) are indicated in panels (a) and (c), respectively, by rectangular boxes with same colors. In all panels, averages are taken over 50 realizations of the phase parameter $\theta$, with $N = 12543$, and $J=1$.
}

    \label{gamma_system_small}
\end{figure}

\begin{figure}
    \centering
    \includegraphics[width=0.48\textwidth]{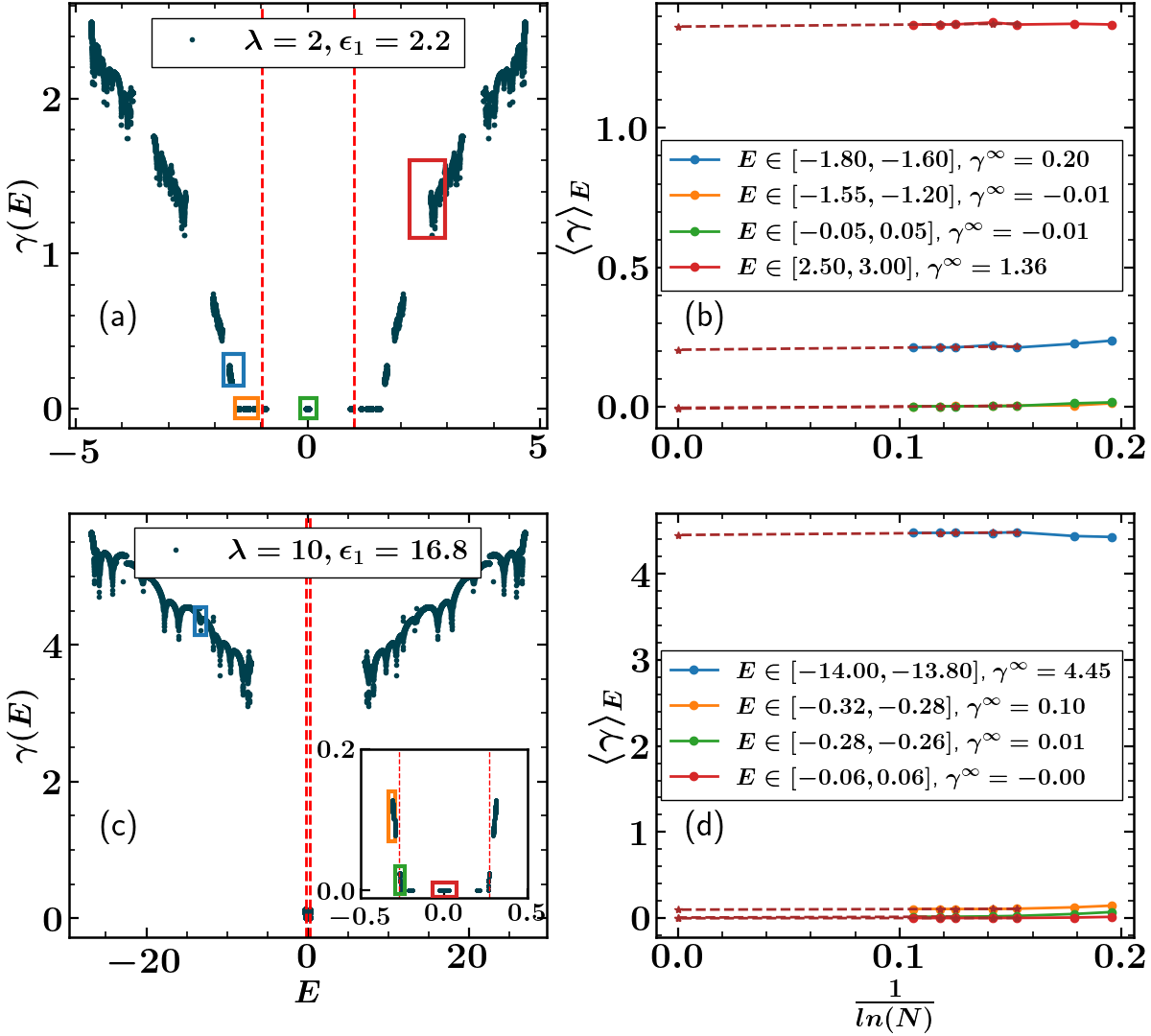}
\caption{Average Lyapunov exponent analysis with system size for $\epsilon_1>\lambda$ regime. For $\lambda=2$, (a) $\gamma(E)$ for all eigenstates with $\epsilon_1 = 2.2$ and (b) average $\langle \gamma \rangle_E$ with $1/\ln N$ in different energy ranges. Panels (c,d) show the same quantities for $\lambda=10$ with $\epsilon_1 = 16.8$. In panel (a) and (c), red dashed lines are the fitted expression Eq.~\eqref{Final_ME_fitting}. Different energy windows used in panels (b) and (d) are indicated in panels (a) and (c), respectively, by rectangular boxes with same colors. In all panels, averages are taken over 50 realizations of the phase parameter $\theta$, with $N = 12543$, and $J=1$.
}

    \label{gamma_system_large}
\end{figure}

\section{Lyapunov exponent with system size}\label{app:Lyapunov_exp_vs_N}

In this section, we study the Lyapunov exponent using exact diagonalization by examining the exponential decay of the wavefunction amplitudes [Eq.~\eqref{Expo_decay_main}], and focus on the Lyapunov exponent as a function of system size, to further confirm the nature of the eigenstates.
To this end, we compute the energy averaged Lyapunov exponent $\langle \gamma \rangle_E$ over different energy intervals corresponding to distinct spectral regions.
We then evaluate $\langle \gamma \rangle_E$ for various system sizes and plot it as a function of $1/\ln(N)$.
The extrapolation of this curve to $1/\ln(N) \to 0$ yields the thermodynamic-limit value $\gamma_{N\to\infty}$.
For localized states, $\gamma_{N\to\infty} > 0$, whereas for ergodic and multifractal states, $\gamma_{N\to\infty} \leq 0$. In addition to the exact diagonalization results, we also computed the Lyapunov exponent directly using the transfer matrix method. The results of this analysis qualitatively agree with those obtained from exact diagonalization, further supporting the robustness of our numerical results.

\begin{figure}
    \centering
    \includegraphics[width=0.48\textwidth]{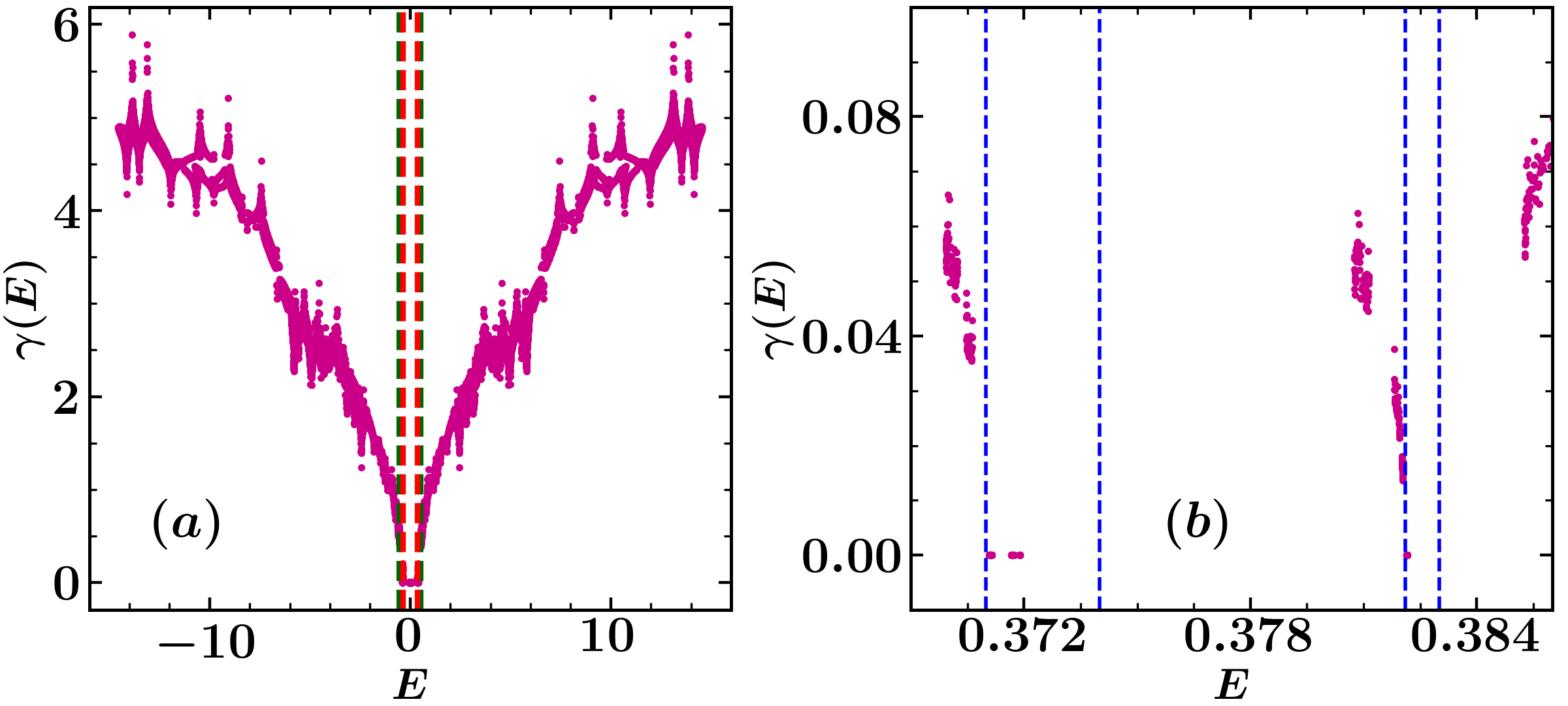}

\caption{
(a) Lyapunov exponent $\gamma(E)$ for all eigenstates with $\epsilon_1 = 4.4$ and $\lambda = 10$.
Green dashed lines denote analytical estimate of the mobility edges Eq.~\eqref{Final_ME}, and red dashed lines correspond to the fitted expression from Eq.~\eqref{Final_ME_fitting}.
(b) Zoomed view of panel (a) highlighting multiple AMEs. The blue dashed lines are guides for eyes drawn to separate
multifractal states from ergodic states.
In all panels, data are averaged over 50 realizations of the phase parameter $\theta$, with $N = 12543$ and $J = 1$.
}

    \label{LE_4.4}
\end{figure}

\begin{figure}
    \centering
    \includegraphics[width=0.48\textwidth]{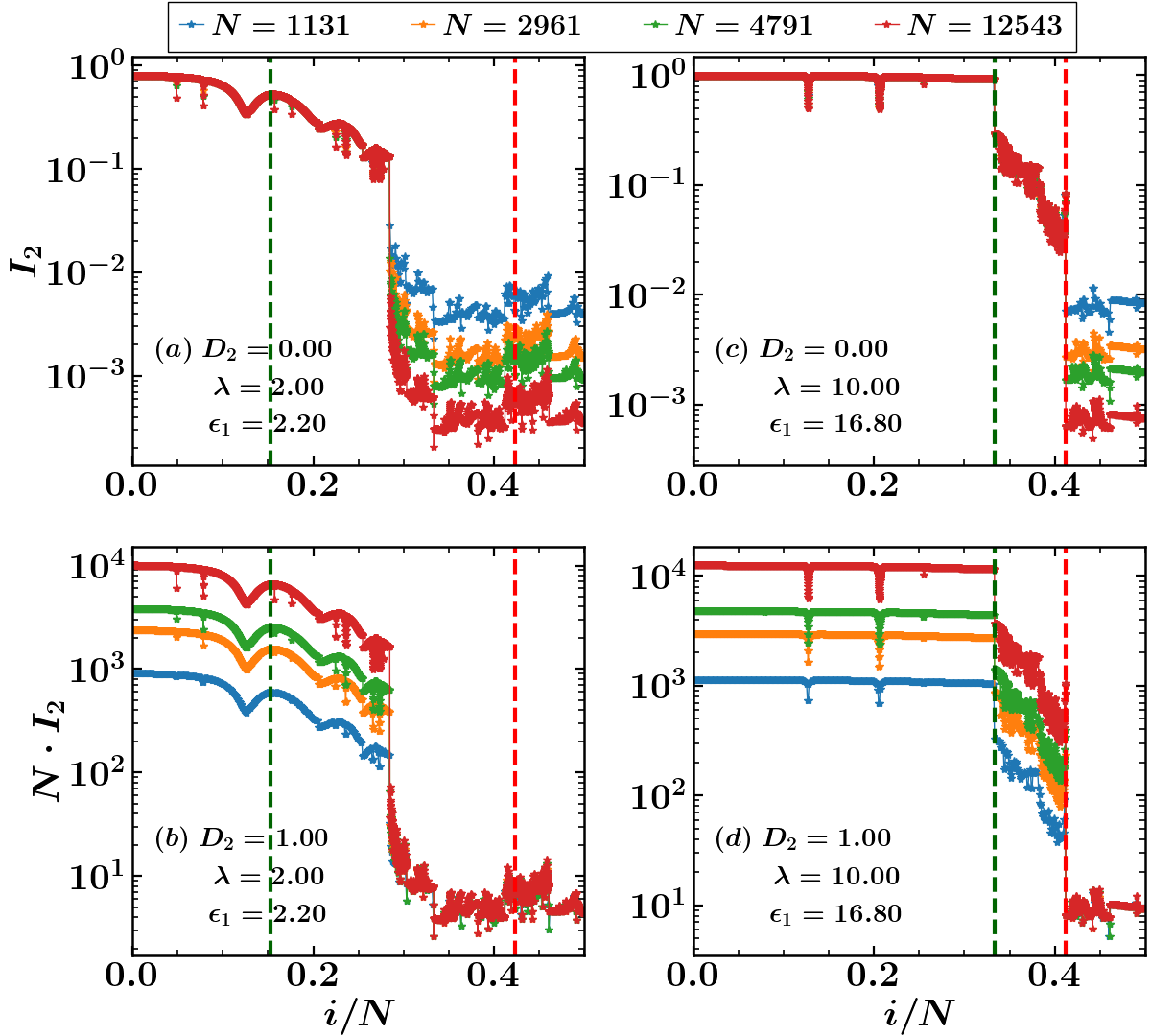}
\caption{IPR collapse plots for different system sizes in $\epsilon_1>\lambda$ regime. The horizontal axis represents the normalized eigenstate index $i/N$. The vertical axis represents $(N^{D_2}I_2)$ for (a)-(b) $\lambda=2$, $\epsilon_1=2.2$, and (c)-(d) $\lambda=10$, $\epsilon_1=16.8$ with $D_2 = 0$ for (a, c) and $D_2 = 1$ for (b, d). The green dashed lines indicate \mk{analytical estimate of the mobility edge, motivated by Avila’s global theory} [Eq. \eqref{Final_ME}]. The red dashed lines are plotted using Eq.~\eqref{Final_ME_fitting} obtained after the fitting procedure. In all panels, averages are taken over 50 realizations of the phase parameter $\theta$. $J=1$.
}

    \label{greater collapse}
\end{figure}

\begin{figure*}
    \centering
    \includegraphics[width=\textwidth]{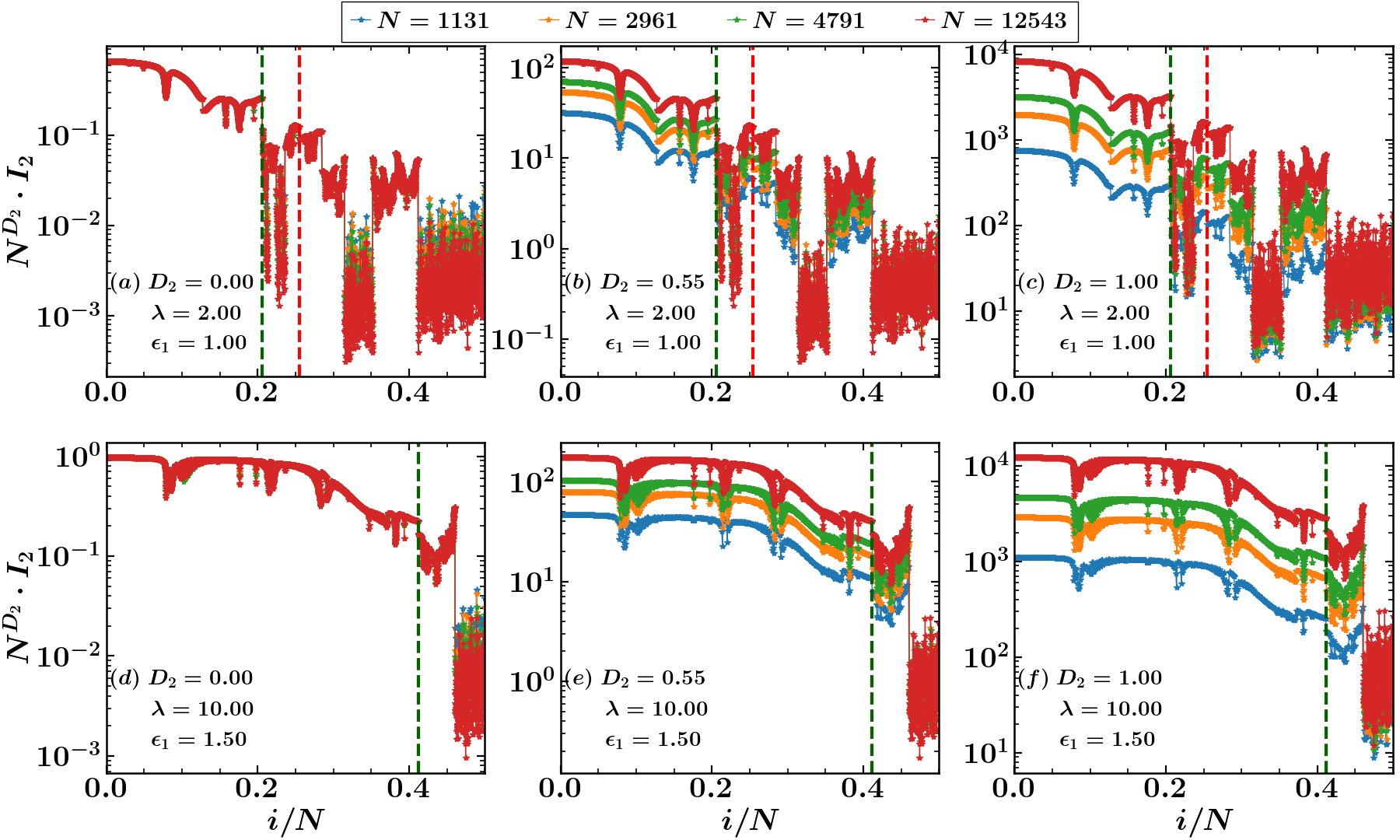}
\caption{IPR collapse plots for different system sizes in the $\epsilon_1<\lambda$ regime. The horizontal axis represents the normalized eigenstate index $i/N$. The vertical axis represents $(N^{D_2}I_2)$ for (a)-(c) $\lambda=2$, $\epsilon_1=1.0$, and (d)-(f) $\lambda=10$, $\epsilon_1=1.50$, with $D_2 = 0$ for (a, d), $D_2 = 0.55$ for (b, e), and $D_2 = 1$ for (c, f). The green dashed lines indicate \mk{analytical estimate of the mobility edge, motivated by Avila’s global theory} [Eq. \eqref{Final_ME}]. The red dashed lines are plotted using Eq.~\eqref{Final_ME_fitting} obtained after the fitting procedure. In all panels, averages are taken over 50 realizations of the phase parameter $\theta$ and $J=1$.
}

    \label{smaller_collapse}
\end{figure*}

\subsection{$\epsilon_1<\lambda$}

We first examine the $\epsilon_1 < \lambda$ regime, where anomalous mobility edges (AMEs) are observed separating localized and multifractal states.
Figures~\ref{gamma_system_small}(a)-(b) show results for $\lambda = 2$ and $\epsilon_1 = 1$.
The Lyapunov exponents are averaged over successive energy intervals from left to right across the spectrum shown by coloured boxes.
A key observation is that $\gamma^\infty \equiv \gamma_{N\to\infty} = -0.02$ ($\approx 0$) for $E \in [-1.50, -1.49]$, indicating multifractal behavior in this range.
Subsequently, $\gamma^\infty = 0.07$ for $E \in [-1.04, -0.98]$, signifying localization, and again $\gamma^\infty \approx 0$ for $E \in [-0.05, 0.05]$, corresponding to multifractal states.
These results clearly demonstrate the presence of multiple AMEs for $\lambda = 2$, a feature not captured by Avila's theory. For comparison, Figs.~\ref{gamma_system_small}(c)-(d) show the results for $\lambda = 10$ and $\epsilon_1 = 1.5$, which again confirm the presence of AMEs in the $\epsilon_1 < \lambda$ regime, consistent with our observations in the main text.

\subsection{$\epsilon_1>\lambda$}
Here, we present a similar analysis in the $\epsilon_1 > \lambda$ regime to confirm the presence of MEs separating localized and ergodic states.
Figures~\ref{gamma_system_large}(a)-(b) show the results for $\lambda = 2$ and $\epsilon_1 = 2.2$, where both $\gamma^\infty \equiv \gamma_{N\to\infty} > 0$ and $\gamma^\infty \approx 0$ are observed, indicating the coexistence of localized and delocalized states.
It is also noteworthy that the analytically\imk{-inspired and fitted estimates of the} MEs (red dashed lines in Fig.~\ref{gamma_system_large}(a)) fail to separate localized and delocalized states, demonstrating the breakdown of the analytical results for $\lambda \lesssim 2$. Furthermore, the presence of MEs is also evident for $\lambda = 10$ and $\epsilon_1 = 16.8$, as shown in Figs.~\ref{gamma_system_large}(c)-(d). These results are consistent with the phase diagrams presented in the main text.

\section{Multiple Mobility edges for $\lambda=10$: $\epsilon_1=4.4 $ case}
\label{app:LE_4.4_multiple}
In this Appendix, we present a representative case with $\lambda = 10$
and $\epsilon_1 = 4.4$, where multiple anomalous mobility edges are
observed. Figure~\ref{LE_4.4}(a) shows the Lyapunov exponent
$\gamma(E)$ calculated for all eigenstates at these parameter
values. As expected, multifractal states appear within the red dashed
AMEs obtained from Eq.~\eqref{Final_ME_fitting}. In addition to these
predicted edges, we identify further multifractal regions in the
narrow energy intervals $0.371 < E < 0.373$ and $0.382 < E < 0.383$,
as shown in Fig.~\ref{LE_4.4}(b). These intervals are marked by blue
dashed lines. Outside these windows, the Lyapunov exponent becomes
finite, indicating localized states. Notably, these additional
multifractal regions lie outside the red dashed AMEs predicted by the
fitted analytical expression. A symmetric set of anomalous mobility
edges is also observed at negative energies, specifically within
$-0.373 < E < -0.371$ and $-0.383 < E < -0.382$. These results
demonstrate that, already at $\lambda = 10$, the system exhibits
multiple anomalous mobility edges, indicating a qualitative deviation
from the predictions of the analytical formalism motivated from
  Avila's theory.

\section{IPR collapse analysis}\label{app:IPR_collapse}
In this section we study scaling of inverse participation ratio with system size to explore the behavior of eigenstates. For a system of size $N$, the IPR scales as $I_2 \sim N^{-D_2}$, where $D_2$ is the fractal dimension. Localized states exhibit $D_2 = 0$, yielding an IPR independent of system size, whereas ergodic states have $D_2 = 1$, for which $I_2 \propto 1/N$. Accordingly, plotting $(N^{D_2} I_2)$ for different system sizes results in a data collapse when the correct $D_2$ is chosen: the curves collapse at $D_2 = 0$ for localized states and at $D_2 = 1$ for ergodic states. For multifractal states to collapse, one can calculate $D_2$  using
\begin{equation}
  D_2=-\frac{\ln\langle I_2\rangle_E}{\ln N}.
\end{equation}
We perform this IPR scaling analysis for multiple system sizes to identify the presence of mobility edges.

\begin{figure*}
    \centering
    \includegraphics[width=\textwidth]{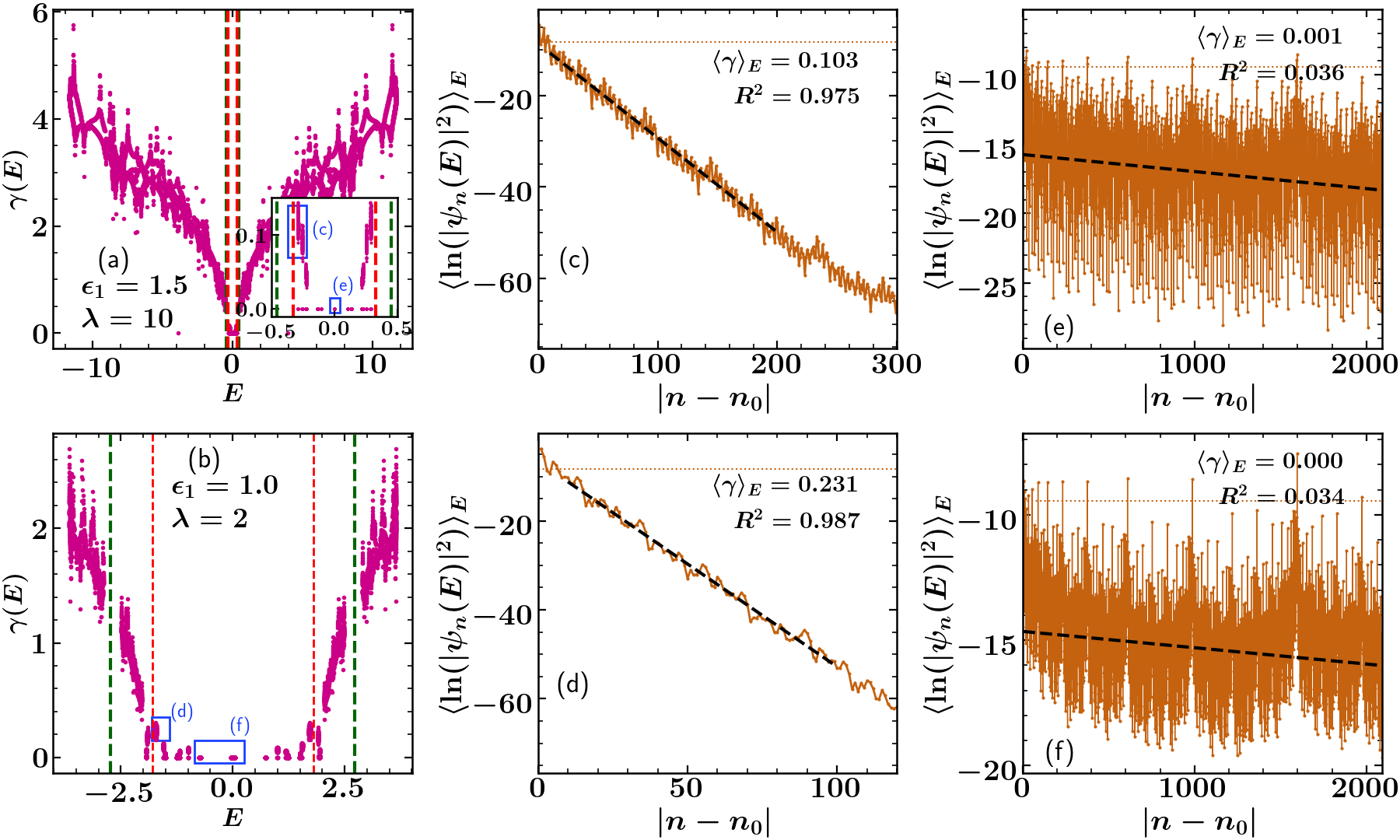}

\caption{
\mk{(a), (b) Lyapunov exponent $\gamma(E)$ across all eigenstates for (a) $\lambda = 10, \epsilon_1 = 1.5$ and (b) $\lambda = 2, \epsilon_1 = 1.0$. Green dashed lines mark the analytical estimate of the mobility edges motivated by Avila's global theory [Eq.~\eqref{Final_ME}], while red dashed lines are plotted using Eq.~\eqref{Final_ME_fitting} obtained after the fitting procedure.
(c)–(f) Averaged $\ln(|\psi_n(E)|^2)$ over eigenstates with energies within the specified intervals, plotted as a function of distance $|n - n_0|$ from the site of maximum amplitude $n_0$.
(c),~(d)~Profiles averaged over (c) $E \in [-0.3, -0.25]$ ($\lambda = 10, \epsilon_1 = 1.5$) and (d) $E \in [-1.8, -1.6]$ ($\lambda = 2, \epsilon_1 = 1.0$). Black dashed lines show exponential fits [Eq. \eqref{Expo_decay_main}] yielding $\langle \gamma\rangle_E = 0.103$ ($R^2 = 0.975$) and $0.231$ ($R^2 = 0.987$), respectively, confirming robust exponential decay and localized states.
(e),~(f)~Profiles averaged over (e) $E \in [-0.004, 0.004]$ ($\lambda = 10, \epsilon_1 = 1.5$) and (f)~$E \in [-0.75, 0.25]$ ($\lambda = 2, \epsilon_1 = 1.0$). Exponential fits yield $\langle \gamma\rangle_E \approx 0$ ($R^2 \approx 0$) in both cases, signifying delocalized behavior.
Blue rectangles in panels (a) and (b) highlight the corresponding energy windows. In all panels, results are shown for a single realization of phase $\theta = 3.5714$, with $N = 12543$ and $J = 1$.}
}

    \label{single_realization}
\end{figure*}

\subsection{$\epsilon_1>\lambda$}

We first examine the regime $\epsilon_1 > \lambda$, where conventional MEs separate localized and ergodic states.
Figures~\ref{greater collapse}(a)-(b) correspond to $\lambda = 2$ and $\epsilon_1 = 2.2$ for $D_2 = 0$ and $D_2 = 1$, respectively.
The data collapse of different system sizes in Fig.~\ref{greater collapse}(a) confirms the localized nature of the states ($D_2 = 0$),
whereas the collapse in Fig.~\ref{greater collapse}(b) indicates ergodic states ($D_2 = 1$).
Only half of the spectrum is shown, as the other half is symmetric. Similarly, Figs.~\ref{greater collapse}(c)-(d) show results for $\lambda = 10$ and $\epsilon_1 = 16.8$, where the same behavior is observed: localized states collapse for $D_2 = 0$ and ergodic states collapse for $D_2 = 1$.
Hence, in this regime, only localized and ergodic states exist. Notably, the expression obtained from Avila-inspired analysis (green dashed lines) fails to quantitatively separate localized and ergodic states, showing only qualitative agreement. In contrast, the fitted expression (red dashed lines) accurately captures the mobility edge for large $\lambda = 10$, but deviates for smaller $\lambda = 2$, consistent with the results presented in the main text. On the other hand, the Avila-inspired analysis, Eq.~\eqref{Final_ME}, separates the strongly localized states with $I_2 \simeq O(1)$ from the ones which are localized on several sites at least for large $\lambda$ values, see Figs.~\ref{greater collapse}(c)-(d). Probably, the former strongly localized states correspond to the slope $\omega(E)=2$, while the latter~--~to smaller, but non vanishing slopes.

\subsection{$\epsilon_1<\lambda$}
Next, we focus on the $\epsilon_1 < \lambda$ regime and present the numerical results of the collapse analysis.
Figures~\ref{smaller_collapse}(a)-(c) correspond to $\lambda = 2$ and $\epsilon_1 = 1$. The data collapse for different system sizes in Fig.~\ref{smaller_collapse}(a) confirms the localized nature of the states. In contrast, Fig.~\ref{smaller_collapse}(c), plotted for $D_2 = 1$, shows no collapse, indicating the absence of ergodic states in this regime. To determine $D_2$ for multifractal states, we plot $\ln \langle I_2\rangle_E$ versus $\ln(N)$ and extract the slope from a linear fit, yielding a fractal dimension $D_2 = 0.54 \pm 0.027$.
Using this value in Fig.~\ref{smaller_collapse}(b), the data collapse confirms the presence of multifractal states.
Moreover, multiple regions in the energy spectrum exhibit multifractal behavior, confirming the existence of multiple mobility edges in the $\epsilon_1 < \lambda$ regime for $\lambda = 2$. Figures~\ref{smaller_collapse}(d)-(f) show similar behavior, further establishing the presence of AMEs separating localized and multifractal states.
Importantly, for $\lambda = 2$, the presence of multiple AMEs indicates that Avila's theory fails even qualitatively, consistent with our findings in the main text. On the other hand, the jumps of the IPR values at the critical energies, given by Eq.~\eqref{Final_ME}, probably separate the strongly localized states, corresponding to the slope $\omega(E)=2$, from the localized ones with smaller, but non vanishing slopes.

\section{Results for single $\theta$ realization}

\label{app:single_realization}
In the main text, all the numerical results are averaged over 50 realizations of the phase parameter $\theta$. To demonstrate that our conclusions do not depend on phase averaging, we present results obtained from a single random realization of $\theta$. For this purpose, we chose a random value $\theta = 3.5714$ within the interval $[0,2\pi]$. Fig.~\ref{single_realization} shows the Lyapunov exponent data for $\lambda = 10$ and $\lambda = 2$ in the regime $\epsilon_1 < \lambda$, corresponding to $\epsilon_1 = 1.5$ and $\epsilon_1 = 1.0$, respectively. For the single realization, the structure of the energy spectrum remains almost unchanged. In particular, the positions of the multiple anomalous mobility edges observed for $\lambda = 2$ [Fig.~\ref{single_realization}(b)] coincide closely with those obtained from the averaged data [Fig.~\ref{spatial_decay_lam_2}(a)]. While minor fluctuations in individual multifractal states are visible, they do not affect the location of the mobility edges.

We further compute the energy averaged Lyapunov exponent $\langle \gamma \rangle_E$ by performing exponential fits over eigenstates within the same energy intervals used in the phase averaged analysis. Figures~\ref{single_realization}(c) and \ref{single_realization}(e) show, respectively, clear exponential decay with $\langle \gamma \rangle_E = 0.103$ ($R^2 = 0.975$) and the absence of exponential decay, with $\langle \gamma \rangle_E = 0.001$ and a poor goodness of fit ($R^2 = 0.036$). This may be compared with the values $\langle \gamma \rangle_E = 0.104$ ($R^2 = 0.975$) and $\langle \gamma \rangle_E = 0.000$ ($R^2 = 0.014$), respectively, for $\lambda = 10$ and $\epsilon_1 = 1.5$ as obtained in the figures \ref{spatial_decay_lam_10}(c) and \ref{spatial_decay_lam_10}(e) where an averaging over 50 realizations of $\theta$ has been carried out. A similarly good agreement is observed for $\lambda = 2$ and $\epsilon_1 = 1.0$, as shown in Figs.~\ref{single_realization}(d)–\ref{single_realization}(f), which are also close  to the corresponding $\theta$-averaged results [see Fig. \ref{spatial_decay_lam_2}(c) and \ref{spatial_decay_lam_2}(e)]. We have verified that this consistency persists across other parameter sets as well. These results confirm that the mobility edge structure and the associated phase characterization are robust and are not an artifact of the $\theta$-averaging over the phase parameter $\theta$.

\bibliography{refs}

\end{document}